\documentstyle[12pt]{article}

\global\arraycolsep=1pt
\begin{document}

\hfill    SISSA/ISAS 147/93/EP

\hfill    hepth@xxx/9309085

\hfill    September, 1993

\begin{center}
\vspace{24pt}
{\large \bf PREDICTIVITY AND NONRENORMALIZABILITY\, \footnotemark
\footnotetext{Partially supported by EEC,
Science Project SC1$^*$-CT92-0789.}
}
\vspace{24pt}

{\sl Damiano Anselmi}

\vspace{6pt}

International School for Advanced Studies (ISAS), via Beirut 2-4,
I-34100 Trieste, Italy\\
and Istituto Nazionale di Fisica Nucleare (INFN) - Sezione di Trieste,
Trieste, Italy\\

\vspace{12pt}

\end{center}

\vspace{24pt}

\begin{center}
{\bf Abstract}
\end{center}

\vspace{12pt}

\noindent
We consider the problem of removing the divergences in an arbitrary gauge-field
theory (possibly nonrenormalizable). We show that
this can be achieved by performing, order by order in the loop expansion,
a redefinition of some parameters (possibly infinitely many)
and a canonical transformation (in the sense of
Batalin and Vilkovisky) of fields and BRS sources.
Gauge-invariance is turned into
a suitable quantum generalization
of BRS-invariance. We define quantum observables and study their properties.
We apply the result to renormalizable gauge-field
theories that are gauge-fixed with a nonrenormalizable gauge-fixing
and prove that their predictivity is retained.
A corollary is that topological field theories are
predictive.
Analogies and differences with the formalisms of
classical and quantum mechanics are pointed out.
\vfill
\eject

\section{Introduction}
\label{intro}

Suppose one is studying an ordinary renormalizable gauge field theory and
that, for some unspecified reason, one wants to choose a
nonrenormalizable gauge-fixing, namely a
gauge-fixing that gives rise to nonrenormalizable vertices in
the BRS action.
In the present paper, we want to study the problem of predictivity of these
theories.
We determine in full generality the algorithm that permits to remove the
divergences of a gauge-field theory,
order by order in the perturbative loop expansion.
We show that the independence of the physical amplitudes
from the continuous deformations of the gauge-fixing survives the
correction algorithm (apart from
eventual BRS anomalies).

The gauge-fixing is thus supposed to contain constants of
negative dimensions in mass units. We denote
them collectively by $\kappa$ throughout
this paper.  The classical renormalizable action ${\cal L}_{class}$
is $\kappa$-independent. One
expects the quantum field theory to retain its predictivity,
although infinitely many types of $\kappa$-dependent
counterterms are needed.
What guarantees the preservation of predictivity
is not {\sl a priori} obvious. The
theorems about renormalizability
of gauge-field theories
that appear in the literature \cite{itzykson,justin,stelle}
are adapted to renormalizable
BRS actions, i.e.\ renormalizable field theories that are gauge-fixed with
an ordinary renormalizable gauge-fixing. Nevertheless, the investigation
of the problem that we have in mind turns out to be very instructive.

Let $\Sigma_0$ be the starting BRS action.
We find it convenient to use the formalism of Batalin and Vilkovisky
\cite{batalin,batalin2,batalin7,batalin3},
although we never distinguish between irreducible and reducible field
theories, to retain generality.
Instead of speaking of antifields $\Phi^*$, we shall speak of
BRS sources $K$.
$\Sigma_0$ has to satisfy the master equation \cite{batalin}.
We make the sufficiently general
assumptions that $\Sigma_0$ is linear in $K$ and that the functional measure
is BRS invariant: $(\Sigma_0,\Sigma_0)=0$ and $\Delta \Sigma_0=0$.
Precisely,
$\Sigma_0$ is the sum of the classical Lagrangian
${\cal L}_{class}(\phi,\lambda)$
plus the BRS variation $s\Psi$ of the gauge-fermion $\Psi(\Phi)$ plus
the terms $K_As\Phi^A$ that couple the BRS sources
$K_A$ to the BRS variations $s\Phi^A$ of the fields $\Phi^A$.
$\lambda$ denote the constants that multiply the gauge-invariant terms
of ${\cal L}_{class}$.
Batalin and Vilkovisky prove \cite{batalin} that
if the action satisfies
the master equation, then the functional integral $Z$
is independent
from the continuous deformations of the gauge-fermion $\Psi$.
Since the nonrenormalizability that we
plan to study only comes from $\Psi$, at first sight one could think that
the argument by Batalin and Vilkovisky is enough to assure predictivity.
Indeed,
the renormalized action $\Sigma$ also satisfies the master equation.
However, the structure of $\Sigma$ is not so simple as the structure of
$\Sigma_0$: one is not even sure to be able to identify a
{\sl renormalized} gauge-fermion inside $\Sigma$.
In conclusion, the problem
that we have in mind is to prove that the subtraction procedure can be
performed while preserving the independence of the physical
amplitudes from the continuous deformations of the gauge-fermion
$\Psi$ that gauge-fixes the {\sl starting} action $\Sigma_0$.

The first step is to find the subtraction algorithm in full
generality (i.e.\ for any gauge-field theory,
eventually a nonrenormalizable one). The algorithm
that we find extends the well-known ones\footnotemark
\footnotetext{See for
example \cite{itzykson,justin} in the case of ordinary Yang-Mills theories and
\cite{stelle} in the case of higher derivative quantum gravity.
The latter is an example of algorithm in which the field redefinitions
are not simply field renormalizations; they are nevertheless
independent of the derivatives of the fields and linearity in the BRS
sources is preserved; in general,
instead, the field redefinitions contain derivative terms and linearity in
the BRS sources is lost, as we shall see
in an example towards the end of the paper.}.

We have
to construct, order by order in the loop expansion,
the renormalized action $\Sigma$ that is able to guarantee
the removal of all the divergences in the effective action $\Gamma$
while retaining gauge-invariance (which is in fact converted into
invariance defined by new nilpotent operators),
although infinitely many
kinds of counterterms are needed.
The correction of divergences is achieved by performing, order by order,
a redefinition of the constants $\lambda$ that multiply the gauge-invariant
terms of the classical Lagrangian ${\cal L}_{class}$,
together with a redefinition of fields
and BRS sources. In particular, the redefinition of fields and BRS sources is a
canonical transformation, in the sense of Batalin and Vilkovisky
\cite{batalin,batalin2,batalin3},
i.e.\ a transformation that preserves the antibrackets
$(\,  . \, , \, . \, )$.
This is the key observation of the paper.

In other words, one starts from a {\sl classical} BRS action $\Sigma_0$
that is the most simple solution to
the master equation, i.e.\ it is linear in the BRS sources (notice that
$\Sigma_0$ is only {\sl one} particular solution to the
master equation).
Then one implements the correction algorithm, which
automatically yields a {\sl quantum} action $\Sigma$,
such that the effective action $\Gamma$ is convergent.
$\Sigma$ is related to $\Sigma_0$ by a set of canonical transformations
and redefinitions
of suitable parameters $\lambda$. Consequently, $\Sigma$ is another
particular solution to the master equation. As a matter
of fact, $\Sigma$ is the {\sl good} solution. The only problem
is that one does not know the good solution $\Sigma$ to the master equation
from the beginning. The subtraction algorithm can thus
be considered as the {\sl principle of
correspondence} that produces the right {\sl quantum}
action $\Sigma$ starting from the known {\sl classical}
action $\Sigma_0$. Such a correspondence principle is nothing but
the search for the {\sl correct} variables $\{\Phi,K\}$ and
the correct definitions of the parameters $\lambda$ of the classical
Lagrangian ${\cal L}_{class}$. One has nonrenormalizability
when the correspondence principle produces infinitely many
good solutions $\Sigma$ and there is no way of privileging
a finite subset of them. Indeed, when a theory is nonrenormalizable,
${\cal L}_{class}$ must depend on infinitely many $\lambda$'s,
otherwise one cannot remove the divergences by {\sl redefining} them.

The second step is the application of
the general subtraction algorithm to renormalizable field theories
that are treated with a nonrenormalizable gauge-fixing.
The fact that divergences can be made disappear
by a set of canonical transformations and redefinitions of parameters
makes the algorithm nicely tractable and all the properties that we need
can be proved without too much difficulty.

The idea is the following: one would like to prove that the infinitely
many $\kappa$-dependent counterterms are all BRS-exact, so that the
normalization of the coupling constants associated to them is
immaterial\footnotemark\footnotetext{As a matter of fact, we are
tacitly assuming that there are not BRS anomalies.
If possible, it is convenient to move the eventual BRS
anomalies from gauge-symmetries
to non-gauge ones and completely neglect the non-gauge symmetries,
otherwise the
master equation cannot be solved.
BRS anomalies can cause a dependence of the physical amplitudes on a coupling
constant that multiplies a BRS-exact term in the Lagrangian
and this can even happen in a
renormalizable theory. Moreover, BRS anomalies can spoil unitarity.
The detailed analysis of these aspects of the problem are beyond the scope
of the present paper and will be eventually treated elsewhere.}.
{\sl A priori} it is not clear what ``BRS-exact counterterm" means, since
the BRS-operator $s$ gets renormalized in a highly nontrivial way
(if it is
compared, for example,
with the simplicity of the renormalization of $s$
when the gauge-fixing is an ordinary renormalizable one).
As a matter of fact, a ``quantum BRS
operator" $\Omega$ can be introduced \cite{omega}: $\Omega$ generalizes $s$ to
the space of fields and BRS sources. It is nilpotent ($\Omega^2=0$)
and is fundamental for the definition of the observables.

We show that, if the $\kappa$-derivative
$\partial \Sigma_0\over \partial \kappa$
of the zeroth loop order BRS action $\Sigma_0$ is $s$-exact
(as it happens when only the gauge-fermion $\Psi$ depends on $\kappa$),
then the derivative $\partial \Sigma\over \partial \kappa$
of the renormalized action $\Sigma$ is $\Omega$-exact and the
derivative $\partial \Gamma\over \partial \kappa$ of the effective action
$\Gamma$ is $(\, . \, ,\Gamma)$-exact. These properties permit to
derive the conclusion that any physical amplitude is $\kappa$-independent.


Regularization is always understood.
The dimensional technique
and the minimal subtraction scheme
can be convenient for most purposes \cite{itzykson,justin,stelle}.
Notice, however, that the operator $\Delta$ introduced by Batalin and
Vilkovisky is proportional to $\delta(0)$ and dimensional regularization
looses any trace of these terms
(see \cite{vanproeyen}
for the study of the Pauli-Villars regularization of $\Delta$
and \cite{anselmi2} for a discussion
of $\delta(0)$ divergent terms related to the measure in quantum gravity).
It is not necessary
to use a regularization scheme that preserves explicit gauge-invariance,
since gauge-invariance can always be retrieved with local counterterms.
The formal arguments of the present paper can suggest
what quantum generalization of gauge-invariance
has to be retrieved with suitable local counterterms
when the regularization scheme explicitly breaks it.

A remark is required in order to specify what we shall mean by
``locality'' (local functional, local canonical transformation, and so on).
Indeed, when a nonrenormalizable Lagrangian is allowed, an arbitrary number
of derivatives can appear. This is not, strictly speaking, a problem
intrinsic in nonrenormalizable theories. It is also present in
renormalizable theories, when one wants to study amplitudes that
involve insertions of composite local operators ${\cal O}(x)$
of sufficiently high dimensionality. In that case, some constant $\kappa$
of negative dimension has to be introduced and infinitely many kinds of
counterterms can appear. What one can do is to truncate each step of the
derivation and each formula up to some order $\kappa^D$. In this way,
all amplitudes with $D$ or less ${\cal O}(x)$-insertions
are made finite. Similarly, we can always assume to neglect the
powers greater than $D$ in the negatively dimensioned
parameters and, contemporarily, to neglect
all the parameters
of dimension less than $-D$.
Then, the effective action is only
made finite up to a given order in the negatively dimensioned parameters
(but to all orders in $\hbar$). A ``local'' functional is a
functional such that its ``$D$-truncation'' is local.
Since the trick can be used for arbitrarily large $D$,
none of the results gets affected.

For simplicity, we can assume that the propagators are of
the form $1\over k^2+m^2$. In other words, the propagator is defined
from the Lagrangian term ${\cal Q}$
that is quadratic both in the fields and in derivatives plus the mass
terms.
All the other terms will be considered as interactions.
This means that in our perturbation series all the parameters but
the constant $\lambda_0$ that multiplies
${\cal Q}$ are considered ``small''.
$\lambda_0$, instead, is considered finite and not small.
This remark is required, since the renormalizablity or nonrenormalizability
of the theory depends on what parameters are considered small
and what are considered finite. In higher derivative quantum gravity
\cite{stelle}, for example, if the constants that multiply
the higher derivative terms $\sqrt{g}R^2$ and
$\sqrt{g}R_{\mu\nu}R^{\mu\nu}$ in the
classical Lagrangian are ``small'', then the theory is nonrenormalizable;
if, on the other hand, they are considered as finite
(on the same footing as
the parameter that multiplies the Hilbert-Einstein term
$\sqrt{g}R$), then the theory is renormalizable (but not unitary).
As a matter of fact, our arguments are also valid if the propagators
behave like $1\over (k^2)^n$,
for $k^2\rightarrow\infty$, as long as $n<\infty$.
In other words, only a finite number of terms that are
quadratic in the fields
can be multiplied by finite (not ``small'') parameters,
otherwise the previous observations about locality are nonsense
(the theory is truly nonlocal).

Let us now make some comments of a general character.
A quantum field theory {\sl can} depend
on infinitely many arbitrary parameters, without
loosing predictivity: it happens when the on-shell physical amplitudes
depend only on finitely many parameters.
If this is the case, the remaining infinitely many parameters
can be fixed at any energy at whatever value one prefers.

This fact suggests that
the following classification of quantum field theories deserves attention.
Let us divide quantum field theories into predictive and
nonpredictive theories.
Predictive field theories are those theories, whose on-shell
physical amplitudes depend on finitely many parameters. Nonpredictive
field theories are those theories, whose on-shell amplitudes depend
on infinitely many parameters.

Renormalizable field theories are
{\sl a fortiori} predictive. Treated with the usual gauge-fixings,
they necessarily depend on finitely many parameters.
However, predictive field theories do not
necessarily depend on finitely many parameters, nevertheless
the physical amplitudes can still depend on
finitely many parameters.
This is for example the case of renormalizable theories
that are treated
with a nonrenormalizable gauge-fixing, on which
we focus in the present paper. Correspondingly, nonrenormalizable
field theories are not necessarily nonpredictive. For
example, finite theories can be nonrenormalizable and predictive.

What are the possible applications of our results?
First of all, notice that a subtraction
algorithm that is applyable to nonrenormalizable field
theories can also be used to study the renormalization of
highly-dimensioned composite operators in ordinary renormalizable field
theories.
Concerning nonrenormalizable gauge-fixings, on the other hand, it is hard
to think such gauge-fixings will ever be used
in Yang-Mills theories or in the Standard Model, since the
ordinary renormalizable ones are quite satisfactory.
However, there are theories in which an exotic gauge-fixing can
be convenient for some peculiar reasons. This is the case, for example,
of four dimensional topological fields theories of Witten type
\cite{witten,4dtopconf,4dtop,4dtopgrav,anselmifre,anselmifresigma},
that are now attracting a lot of interest. Among these, an
interesting one is surely the topological sigma-model formulated in
\cite{anselmifresigma}, which describes the triholomorphic embeddings
of four dimensional Riemannian manifolds into
almost quaternionic manifolds\footnotemark\footnotetext{
I would like to thank E.\ Witten for a couple of
inspiring discussions on the nonrenormalizability of this model,
that gave rise to my interest in the problem investigated in the present
paper.}.
It is an example of an irreducible gauge-field theory, which
is nonrenormalizable, but, since it is also topological,
the nonrenormalizability is entirely due to the gauge-fixing
(the classical action is either zero or a topological invariant).
It is well-known that the mathematical interpretation of
topological field theories shows that there is a
dependence on the gauge-fixing: two gauge-fixings that
are not continuously deformable one into the other give rise to
inequivalent field theories\footnotemark\footnotetext{A similar
behaviour, of course, is expected to happen in {\sl any}
gauge-field theory.}. Thus one cannot
turn the gauge-fixing (which in the present case is
the triholomorphicity condition on the map) to a renormalizable one
without either
spoiling general covariance or changing completely the theory.
Nevertheless, one expects the nonrenormalizable
theory to be perturbatively well defined
and predictive. This is a corollary of our result,
namely topological field theories (of Witten type) are predictive.

Notice that topological field theories
are intrinsically nonperturbative. On the other hand, in
this paper we are only concerned with the perturbative behaviour
of quantum field theories. Our results can thus by described by saying
that the
nonrenormalizability of topological models
produces no perturbative obstruction to a good definition of them
(apart from the eventual BRS anomalies).

An example
of first stage reducible nonrenormalizable topological gauge-field
theory is topological gravity (there are various versions of it:
topological conformal gravity, see \cite{4dtopconf};
topological gravity with the self-duality condition on the Riemann
tensor \cite{4dtop}; topological gravity derived from
the twist of N=2 supergravity \cite{4dtopgrav}).
The topological Yang-Mills theory originally formulated by Witten
\cite{witten} is, instead, renormalizable, so that its
perturbative definition is straightforward.

An exotic gauge-fixing can also be used in order to make computations easier.
For example, in ref.\ \cite{vandeven} a nonlinear gauge-fixing plays an
essential role in simplifying the perturbative two-loop
computation in quantum gravity\footnotemark\footnotetext{That
computation
confirmed the Goroff-Sagnotti result \cite{sagnotti} that
quantum gravity is not two-loop finite.}.

The paper is organized as follows. In section \ref{preliminars} we fix
notation and conventions and give the fundamental properties
that will be useful in successive derivations. In section \ref{observables}
we define the nilpotent quantum operators $\Omega$
and ${\rm ad}\, \Gamma$ that generalize
the BRS operator $s$. We define the observables and study their
change under canonical transformations. We point out analogies and
differences with the formalisms of classical and quantum mechanics.
In section \ref{nonrenormalizable} we derive the algorithm for removing
the divergences of a generic quantum field theory, while preserving
gauge-invariance. In section \ref{predictivity} we show that the
redefinitions of the parameters $\lambda$ of ${\cal L}_{class}$
do not depend on the constants that are only introduced
{\sl via} the gauge-fixing. This result applies
to the case of renormalizable gauge-field theories treated with a
nonrenormalizable gauge-fixing showing that predictivity is retained.
We also define the convergent physical amplitudes. In section
\ref{examples} we give some examples: Q.E.D. with an exotic gauge-fixing
and the topological $\sigma$-model of ref.\ \cite{anselmifresigma}.
We also show how the usual coupling constant renormalizations
and wave function renormalizations are retrieved within our approach.
Section \ref{conclusions} contains the conclusions, while
the appendix is devoted to the lengthy, but straightforward proof
of a formula that is needed in the paper.

\section{Preliminars}
\label{preliminars}

In this section we introduce the notation and the basic definitions. For
self-consistence, we also report some simple arguments by
Batalin and Vilkovisky \cite{batalin,batalin2,batalin3}
that will be useful in the following.

The partition function is
\begin{equation}
Z[J_A,K_A]=\int  {\rm d}\Phi\, {\rm e}^{{i\over \hbar}
\Sigma(\Phi^A,K_A)+{i\over \hbar}J_A\Phi^A},
\label{partitionfunction}
\end{equation}
where $\Phi^A$ denote the fields, while
$J_A$ are the corresponding sources.
Notice that $\Phi^A$ is not the {\sl minimal} set of fields \cite{batalin}
that is usually
necessary to solve the master equation. Rather, it is enlarged to
contain the classical fields $\phi$, the ghosts, the antighosts,
the Lagrange multipliers, the eventual extraghosts and so on.
$K_A$ is the source associated
to the BRS transformation $s\Phi^A$ of the field $\Phi^A$
(it will be called the {\sl BRS source}). $K_A$ differs from
the antifields $\Phi^*_A$ introduced by Batalin and Vilkovisky by
a derivative of the gauge-fermion $\Psi$:
\begin{equation}
K_A=\Phi^*_A-{\partial \Psi\over \partial \Phi^A}.
\label{antifields}
\end{equation}
The transition from $\{\Phi^A,\Phi^*_A\}$ to $\{\Phi^A,K_A\}$ is a canonical
transformation \cite{batalin,batalin2}.
The operators that we use are
defined in terms of
$\{\Phi^A,K_A\}$ rather than $\{\Phi^A,\Phi^*_A\}$.
The antibrackets are
\begin{equation}
(X,Y)={\partial_r X\over \partial \Phi^A}{\partial_l Y\over \partial K_A}-
{\partial_r X\over \partial K_A}{\partial_l Y\over \partial \Phi^A}.
\end{equation}
The subscripts $r$ and $l$ denote right and left derivatives,
respectively. When there is no subscript, that means that left and right
derivatives are equivalent.
The statistics of the field $\Phi^A$ is denoted by $\varepsilon_A$, which
is an integer modulo two (zero for bosons, one for fermions).
The statistics of $K_A$ is $\varepsilon_A+1$.
We report some simple properties of the antibrackets \cite{batalin3}
that will be very useful
in the calculations, namely
\begin{eqnarray}
\matrix{\varepsilon[(X,Y)]=\varepsilon(X)+\varepsilon(Y)+1, \cr
(X,Y)=-(-1)^{(\varepsilon(X)+1)(\varepsilon(Y)+1)}(Y,X),\cr
(-1)^{(\varepsilon(X)+1)(\varepsilon(W)+1)}(X,(Y,W))+\,{\rm
cyclic \,\,permutations}\,=0.}
\label{antibracketproperties}
\end{eqnarray}
We also introduce the operator $\Delta$, the definition of which differs from
the one
given by Batalin and Vilkovisky because of the replacements of antifields
with BRS sources, namely
\begin{equation}
\Delta={\partial _r\over \partial \Phi^A}{\partial_r\over \partial K_A}
(-1)^{\varepsilon_A+1}.
\end{equation}
The properties of $\Delta$ that will be useful in the calculations are
$\varepsilon(\Delta)=1$ and
\begin{equation}
\matrix{\Delta^2=0,&\Delta (X,Y)=(X,\Delta Y)-(-1)^{\varepsilon(Y)}
(\Delta X,Y).}
\label{deltaproperties}
\end{equation}

The action $\Sigma$ is supposed to satisfy the master equation
\begin{equation}
(\Sigma,\Sigma)=2i\hbar\Delta\Sigma.
\label{masterequation}
\end{equation}

A canonical transformation is a transformation of fields $\Phi^A$
and BRS sources $K_A$ into new fields $\Phi^{\prime A}$ and new
BRS sources
$K_A^\prime$
that preserves the antibrackets. As in classical mechanics, a generating
functional $F(\Phi^A,K^\prime_A)$ can be introduced, such that
\begin{equation}
\matrix{
\Phi^{\prime A}={\partial F\over \partial K^\prime_A},
&
K_A={\partial F\over \partial \Phi^A}.
}
\label{cantrasf}
\end{equation}
$F$ is a fermionic functional.
The rule for the change of the action under a canonical transformation
${\cal C}$
is determined by the requirement that the new action $\Sigma^\prime$
satisfies the
master equation. Our convention is that the arguments of $\Sigma^\prime$
are still called $\{\Phi^A,K_A\}$. The expression for $\Sigma^\prime
(\Phi^A,K_A)$, that will be proved in a moment, then turns out to be
\begin{equation}
\Sigma^\prime(\Phi^A,K_A)={\cal C}\Sigma(\Phi^A,K_A)=
\Sigma(\Phi^{\prime A}(\Phi,K),
K^\prime_A(\Phi,K))+{1\over 2}i\hbar\ln\,J,
\label{transf}
\end{equation}
where $J$ is the Berezinian determinant associated to the change
of fields and sources (which is not a change of variables in the functional
integral)
\begin{equation}
J=\det{\partial (\Phi,K)\over \partial (\Phi^\prime,K^\prime)}.
\end{equation}
A useful property of $J$ is the following \cite{batalin2}
\begin{equation}
\Delta\, \ln \, J={1\over 4}(\ln \, J , \ln \, J).
\label{deltaj}
\end{equation}

We adopt the following convention: any
$\{\Phi^A,K_A\}$-dependent functional or operator
will be marked with a tilde when we mean that the
fields $\Phi^A$ and the BRS sources $K_A$ have to be replaced by
$\Phi^{\prime A}(\Phi,K)$ and $K_A^\prime (\Phi,K)$, respectively. Thus,
equation (\ref{transf}) will be briefly written as
\begin{equation}
\Sigma^\prime=\tilde\Sigma+{1\over 2}i\hbar\ln\,J.
\label{sigmaprimo}
\end{equation}
Similarly, $\tilde \Delta$ is the same as the operator $\Delta$, in which
the derivatives with respect to unprimed fields and BRS sources are replaced
by derivatives with respect to the primed ones.
The fact that the transformation that we are considering is canonical can
be simply expressed by the following equation
\begin{equation}
(\, . \,  , \, . \, )\,\tilde{}=(\, . \,  , \, . \, ).
\label{canonical}
\end{equation}
The proof that $\Sigma^\prime$ still satisfies the master equation if
$\Sigma$ does is immediate consequence of the following identity
\cite{batalin2}
\begin{equation}
\tilde\Delta X=\Delta X - {1\over 2} (X,\ln \, J),
\label{deltatilde}
\end{equation}
for any $X$. Indeed, let us consider the master equation
(\ref{masterequation}) satisfied by $\Sigma$.
Since the name that one gives to fields and BRS sources is immaterial, it
is clear that the identity
\begin{equation}
(\tilde\Sigma,\tilde\Sigma)\,\tilde{}=2i\hbar\tilde\Delta \tilde\Sigma,
\label{utility}
\end{equation}
also holds.
Using (\ref{canonical}) and (\ref{deltatilde}) one has
\begin{equation}
(\tilde\Sigma,\tilde\Sigma)=2i\hbar\Delta \tilde\Sigma
-i\hbar(\tilde\Sigma,\ln\, J).
\label{utility2}
\end{equation}
At this point, using (\ref{deltaj}),
it is easy to see that definition (\ref{sigmaprimo}) is chosen
precisely to have
\begin{equation}
(\Sigma^\prime,\Sigma^\prime)=2i\hbar\Delta \Sigma^\prime,
\label{utility3}
\end{equation}
as desired.

Now, consider the functional integral (\ref{partitionfunction}).
Let us perform the following infinitesimal change of variables
\begin{equation}
\Phi^A\rightarrow \Phi^A+{\partial _l \Sigma\over \partial K_A}\Lambda
=\Phi^A+(\Phi^A,\Sigma)\Lambda,
\label{chofvar}
\end{equation}
where $\Lambda$ is some constant, infinitesimal fermionic parameter.
Due to the eventual nonlinearity of $\Sigma$ in $K_A$,
(\ref{chofvar}) is in general a source-dependent change of variables.
The variation $\delta Z$ of $Z$ is zero and can be written in the form
\begin{equation}
0=\delta Z=\int {\rm d}\Phi\,
{\rm e}^{{i\over \hbar}\Sigma+{i\over \hbar}J_A\Phi^A}\left\{-{i\over 2 \hbar}
(2i\hbar\Delta\Sigma-(\Sigma,\Sigma))+{i\over \hbar}J_A(\Phi^A,\Sigma)\right\}
\Lambda.
\label{2.19}
\end{equation}
The same result can be derived from the identity
\begin{equation}
0=\int {\rm d}\Phi\, {\partial_r\over \partial \Phi^A}\left\{
{\rm e}^{{i\over \hbar}\Sigma+{i\over \hbar}J_A\Phi^A}\,
{\partial_l\Sigma\over \partial K_A}\right\}.
\end{equation}

Since $\Sigma$ satisfies the master equation (\ref{masterequation}),
formula (\ref{2.19}) reduces to the following, fundamental Ward identity:
\begin{equation}
<J_A(\Phi^A,\Sigma)>_J=0.
\label{ward}
\end{equation}
The subscript $J$ is to mean that the sources $J$ are not set to zero.

Let us introduce, as usual, the generating functional $W[J_A,K_A]$
of connected Green functions, as follows:
\begin{equation}
Z[J_A,K_A]={\rm e}^{{i\over \hbar}W[J_A,K_A]}.
\end{equation}
The Ward identity (\ref{ward}) can be rewritten as
\begin{equation}
J_A{\partial _l W\over \partial K_A}=0.
\label{ward2}
\end{equation}
Let us also introduce
the generating functional $\Gamma[\Phi^A,K_A]$ of one particle
irreducible Green functions, defined as the Legendre transform
of $W[J_A,K_A]$ with respect to $J_A$ and with $K_A$ inert:
\begin{equation}
\Gamma[\Phi^A,K_A]=W[J_A(\Phi,K),K_A]-J_A(\Phi,K)\Phi^A,
\label{gammadefinition}
\end{equation}
where the function $J_A(\Phi,K)$ is defined as the inverse of
\begin{equation}
\Phi^A(J,K)={\partial_l W\over \partial J_A}.
\end{equation}
The properties of Legendre transforms guarantee that
\begin{equation}
\matrix{J_A=-{\partial_r \Gamma\over \partial \Phi^A},&
{\partial_l W\over \partial K_A}={\partial_l \Gamma\over \partial K_A},}
\end{equation}
since the BRS sources $K$ are simple spectators.
Thus the Ward identity (\ref{ward2}) can be rewritten as
\cite{batalin3}
\begin{equation}
(\Gamma,\Gamma)=0.
\label{wardg}
\end{equation}
So, whenever the action $\Sigma$ satisfies the master equation
(\ref{masterequation}), then the effective action $\Gamma$ satisfies
the Ward identity (\ref{wardg}).

\section{Observables}
\label{observables}

In this section, we identify the
observables and study their properties.
Let us first make a brief digression on classical and quantum
mechanics.

In classical mechanics, the Hamiltonian $H(p,q)$ (let us assume it is
time independent) satisfies the ``master equation"
\begin{equation}
\{H,H\}=0,
\label{classicalmaster}
\end{equation}
$\{\, . \, , \, . \, \}$ denoting the Poisson brackets.
A time independent integral of motion
${\cal O}$ satisfies
\begin{equation}
\{{\cal O},H\}\equiv({\rm ad}\, H )\, {\cal O}=0.
\end{equation}
Notice that the operator ${\rm ad}\, H$ is nilpotent, $({\rm ad}\, H)^2=0$.
Suppose $H$ depends on some parameter $g$: $H=H(p,q,g)$.
Consider a time independent (but possibly $g$-dependent)
canonical transformation generated by $f(p^\prime,q,g)$, namely
\begin{equation}
\matrix{p={\partial f\over \partial q},&
q^\prime={\partial f\over \partial p^\prime}.}
\end{equation}
The Hamiltonian $H$ transforms into
\begin{equation}
H^\prime(p,q,g)=H(p^\prime(p,q,g),q^\prime(p,q,g),
g)=\tilde H.
\label{h1}
\end{equation}
Similarly, an integral of motion ${\cal O}$ transforms into ${\cal O}^\prime=
\tilde{\cal O}$.
Since canonical transformations preserve the Poisson brackets,
${\rm ad}\, H$-closure and ${\rm ad}\, H$-exactness are converted into
${\rm ad}\, H^\prime$-closure and ${\rm ad}\, H^\prime$-exactness,
respectively.
We are interested in the transformation of $\partial H\over \partial g$,
i.e.\ $\partial H^\prime \over \partial g$. The explicit computation
gives
\begin{equation}
{\partial H^\prime\over \partial g}=
\widetilde{\partial H\over \partial g}-\left\{{\partial f\over \partial
g},H^\prime\right\}.
\label{de1}
\end{equation}
In particular, this assures that if ${\partial H\over \partial g}$ is
${\rm ad}\, H$-exact, then ${\partial H^\prime\over \partial g}$
is ${\rm ad}\, H^\prime$-exact.

Let us now turn to quantum mechanics. The ``master equation" is simply
\begin{equation}
[H,H]=0,
\label{quantummaster}
\end{equation}
where the square brackets denote the commutator. A (time independent)
observable is an operator ${\cal O}$ that commutes with $H$. A canonical
transformation is performed by a unitary operator $U$ and
\begin{equation}
H^\prime=UHU^{-1}=\tilde H.
\label{h2}
\end{equation}
Similarly, ${\cal O}$ transforms into ${\cal O}^\prime=
\tilde{\cal O}$.
Again, ${\rm ad}\, H$-closure and ${\rm ad}\, H$-exactness are converted into
${\rm ad}\, H^\prime$-closure and ${\rm ad}\, H^\prime$-exactness.
Moreover,
\begin{equation}
{\partial H^\prime \over \partial g}=U{\partial H\over \partial g}
U^{-1}+\left[{\partial U\over \partial g}U^{-1},UHU^{-1}\right]=
\widetilde{{\partial H\over \partial g}}+
\left[{\partial U\over \partial g}U^{-1},H^\prime\right].
\label{de2}
\end{equation}
Once more it is true that if  ${\partial H\over \partial g}$ is
${\rm ad}\, H$-exact, then ${\partial H^\prime\over \partial g}$
is ${\rm ad}\, H^\prime$-exact.

Inspired by this digression, it is simple to work out the definition of
the nilpotent operator
$\Omega$ that extends the BRS operator $s$ to the space of fields
and BRS sources \cite{omega}. This will also be useful to understand the limits
of the analogy among antibrackets, commutators and Poisson brackets.
Suppose $\Sigma$ depends on some parameter $g$.
The change of ${\partial \Sigma\over \partial g}$ under
a canonical transformation (\ref{cantrasf}) generated by
$F(\Phi^A,K_A^\prime,g)$
turns out to be very similar to (\ref{de1}) and (\ref{de2}), namely
\begin{equation}
{\partial\Sigma^\prime\over \partial g}=
\widetilde{\partial \Sigma\over \partial g}-\Omega^\prime
{\partial F\over \partial g}
\label{SIGMAPRIMOSULAMBDA}
\end{equation}
(the proof can be found in the appendix),
where we have introduced the fundamental operator
\begin{equation}
\Omega^\prime={\rm ad}\,\Sigma^\prime-i\hbar\Delta,
\end{equation}
${\rm ad}\, \Sigma$ being now defined by $({\rm ad}\, \Sigma)\, X=(X,\Sigma)$.
We are thus lead to interpret the operator $\Omega^\prime$
as the canonically
transformed version of the operator $\Omega$, defined by
\begin{equation}
\Omega X\equiv (X,\Sigma)-i\hbar\Delta X.
\end{equation}
Notice that $\Omega$ acts on the space of fields and BRS sources.
$\Omega$ is the candidate for the
definition of the physical quantities.
The key check that we are on the right way is the nilpotence of $\Omega$.
Indeed, let us suppose that $X=\Omega \chi$ for some $\chi$.
Then,
\begin{equation}
\Omega X=\Omega^2 \chi=((\chi,\Sigma)-i\hbar\Delta\chi,\Sigma)
-i\hbar\Delta ((\chi,\Sigma)-i\hbar\Delta \chi).
\end{equation}
Using the Jacobi identity of (\ref{antibracketproperties})
and the properties (\ref{deltaproperties})
of the operator $\Delta$, we get
\begin{equation}
\Omega^2\chi={1\over 2}(\chi,(\Sigma,\Sigma))-i\hbar(\chi,\Delta \Sigma),
\end{equation}
which vanishes due to the master equation (\ref{masterequation}).

Differentiation of the master equation (\ref{masterequation}) gives
\begin{equation}
\left({\partial\Sigma\over \partial g},\Sigma \right)=
i\hbar\Delta {\partial\Sigma\over \partial g}.
\end{equation}
This means that ${\partial\Sigma\over \partial g}$ is $\Omega$-closed.
We then expect ${\partial\Sigma^\prime
\over \partial g}$ to be $\Omega^\prime$-closed.
In particular, as we shall prove in a moment,
the first term on the right hand side of
equation (\ref{SIGMAPRIMOSULAMBDA}) is $\Omega^\prime$-closed,
while the second term is trivially $\Omega^\prime$-exact.
In classical or quantum mechanics,
we cannot say that ${\partial H\over \partial g}$ is
${\rm ad }\, H$-closed: indeed, a $g$-differentiation of
equations (\ref{classicalmaster}) and (\ref{quantummaster})
gives no new information.
This is because (\ref{classicalmaster}) and
(\ref{quantummaster}) are trivial identities, i.e.\
they would be satisfied by any $H$. The master equation
(\ref{masterequation}), however, has a nontrivial content: the definition
of antibrackets does not imply that it is identically
true\footnotemark\footnotetext{As a matter of fact,
the definition of antibrackets only implies that $(F,F)=0$ for any fermionic
$F$.}. This suggests that the analogy with classical and quantum
mechanics has to be taken {\sl cum grano salis}.
This is a luck, rather that a handicap, since we shall be able
to prove useful properties that have
no counterpart in classical or quantum mechanics. In particular,
the nontriviality of
(\ref{wardg}) will be extremely
important in the following.

The proof of formula (\ref{SIGMAPRIMOSULAMBDA})
is a little involved, although conceptually rather simple
[it is only a matter of change of variables and it
can be obtained  by mimicking the analogous steps that permit to prove
(\ref{de1}) in classical mechanics]. That is why we
postpone it to the appendix.
For the moment, the reader should be satisfied with
the plausibility that is suggested by the similarities
with equations (\ref{de1}) and (\ref{de2}) and
the fact that $\Omega^\prime$ is the only reasonable nilpotent
generalization of the BRS operator $s$ to the space of fields and BRS
sources.
The minus sign in (\ref{SIGMAPRIMOSULAMBDA})
can be promptly checked by considering a canonical
transformation infinitesimally close to the identity,
i.e.\ a transformation with $F(\Phi^A,K_A^\prime)=
\Phi^AK_A^\prime+\varepsilon {\cal R}(\Phi^A,K_A^\prime)$,
$\varepsilon$ being a constant infinitesimal parameter.

Let us prove in detail that if ${\cal O}$ is $\Omega$-closed,
then $\tilde{\cal O}$ is $\Omega^\prime$-closed. In particular,  this
assures that $\widetilde{\partial \Sigma\over \partial g}$ is
$\Omega^\prime$-closed. Indeed, we have
\begin{equation}
\Omega {\cal O}=({\cal O},\Sigma)-i\hbar\Delta {\cal O}=0.
\end{equation}
Changing name to fields and BRS sources wherever their appear, we also get
\begin{equation}
(\tilde{\cal O},\tilde\Sigma)\,\tilde{}-i\hbar\tilde\Delta \tilde{\cal O}=0.
\end{equation}
Using (\ref{sigmaprimo}), (\ref{canonical}) and (\ref{deltatilde}), we get
\begin{equation}
(\tilde{\cal O},\Sigma^\prime)-i\hbar\Delta \tilde{\cal O}=
\Omega^\prime \tilde{\cal O}=0.
\end{equation}
Thus, if $\cal O$ is $\Omega$-closed, it is natural
to define its variation under a canonical transformation
according to
\begin{equation}
{\cal O}^\prime=\tilde{\cal O}+\Omega^\prime \Lambda.
\label{observabletransform}
\end{equation}
In general, $\Lambda$ can be chosen to be zero. Notice, however,
that for ${\cal O}={\partial \Sigma\over \partial g}$ one has
$\Lambda=-{\partial F\over \partial g}\neq 0$,
according to (\ref{SIGMAPRIMOSULAMBDA}).
With $\Lambda=0$, (\ref{observabletransform}) is
reminiscent of the usual quantum mechanical rule
${\cal O}^\prime=U{\cal O}U^{-1}=\tilde{\cal O}$.

We now prove that if ${\cal O}$ is $\Omega$-exact,
then $\tilde{\cal O}$ is $\Omega^\prime$-exact.
Let $\chi$ be such that ${\cal O}=\Omega \chi$. We have
\begin{equation}
{\cal O}=(\chi,\Sigma)-i\hbar\Delta \chi.
\end{equation}
Changing name to fields and BRS sources wherever their appear, we also get
\begin{equation}
\tilde {\cal O}=(\tilde\chi,\tilde\Sigma)\,\tilde{}-
i\hbar\tilde\Delta \tilde\chi.
\end{equation}
Using (\ref{sigmaprimo}), (\ref{canonical}) and (\ref{deltatilde}), we
finally obtain
\begin{equation}
\tilde{\cal O}=(\tilde\chi,\Sigma^\prime)-i\hbar\Delta \tilde\chi=
\Omega^\prime \tilde\chi,
\end{equation}
as desired. Consequently, ${\cal O}^\prime$ is also
$\Omega^\prime$-exact. A corollary is that if ${\partial \Sigma\over
\partial g}$ is $\Omega$-exact, then ${\partial \Sigma^\prime\over
\partial g}$ is $\Omega^\prime$-exact.
This fact will be very useful in the following.

The formul\ae\ for the change of the functional $\Gamma$ under
a canonical transformation can be also conjectured by analogy
with the classical and quantum mechanical formul\ae.
In particular, (\ref{h1}) and (\ref{h2})
suggest
\begin{equation}
\Gamma^\prime=\tilde\Gamma.
\label{gammaprimo}
\end{equation}
Moreover, (\ref{de1}) and (\ref{de2}) suggest
\begin{equation}
{\partial \Gamma^\prime\over \partial
g}=\widetilde{\partial \Gamma\over \partial g}-
\left({\partial F\over\partial g},\Gamma^\prime\right)
\equiv \widetilde{\partial \Gamma\over \partial g}-
{\rm ad}\, \Gamma^\prime\,{\partial F\over\partial g}.
\label{gammaprimosulambda}
\end{equation}
Although formula (\ref{gammaprimo}) is very natural and can be easily
checked in the case of canonical transformations infinitesimally
close to the identity, we give it only as a conjecture.
We shall be able to prove our results independently of (\ref{gammaprimo}).
Equation (\ref{gammaprimosulambda}), instead,
can be simply proved by taking the $g$-derivative
of (\ref{gammaprimo}) and mimicking the standard arguments that
are contained in the appendix.
However, it must be kept in mind that it depends on the
validity of (\ref{gammaprimo}), that we leave without proof.
A direct consequence of (\ref{gammaprimosulambda})
is that if ${\partial \Gamma\over \partial
g}$ is ${\rm ad}\,\Gamma$-exact, then
${\partial \Gamma^\prime\over \partial
g}$ is ${\rm ad}\, \Gamma^\prime$-exact.

Moreover, a $g$-differentiation of (\ref{wardg}) shows that $\partial\Gamma
\over\partial g$ is ${\rm ad}\, \Gamma$-closed.

Let us now prove that
if ${\cal O}$ is $\Omega$-closed, then $<{\cal O}>_J$
is ${\rm ad}\,\Gamma$-closed
(the subscript $J$ means that the sources $J$ are not set to zero).
We start from
\begin{equation}
0=<\Omega {\cal O}>_J=<
{\partial_r {\cal O}\over \partial \Phi^A}{\partial_l\Sigma\over
\partial K_A}-{\partial_r {\cal O}\over \partial K_A}
{\partial_l\Sigma\over
\partial \Phi^A}-i\hbar (-1)^{\varepsilon_A+1}{\partial_r\over \partial\Phi^A}
{\partial_r {\cal O}\over \partial K_A}>_J.
\label{gammao}
\end{equation}
By performing the change of variables
\begin{equation}
\delta\Phi^A={\partial _l\Sigma\over \partial K_A}\Lambda,
\label{chofv}
\end{equation}
($\Lambda$ being a fermionic infinitesimal parameter) in the functional
integral $<{\cal O}>_J$,
one finds
\begin{equation}
<{\partial_r {\cal O}\over \partial \Phi^A}{\partial_l\Sigma\over
\partial K_A}>_J=-{i\over \hbar}<{\cal O}
{\partial_l\Sigma\over
\partial K_A}>_J\, J_A.
\label{pr1}
\end{equation}
Moreover, with an integration by parts one can show that
\begin{equation}
<\Delta {\cal O}>_J={i\over \hbar}
<{\partial_r {\cal O}\over \partial K_A}
{\partial_l\Sigma\over
\partial \Phi^A}>_J+{i\over \hbar}<{\partial_r{\cal O}\over \partial K_A}>_J
\, J_A (-1)^{\varepsilon_A}.
\label{pr2}
\end{equation}
It is thus possible to write (\ref{gammao}) in the form
\begin{eqnarray}
0&=&
{i\over \hbar}<{\cal O}{\partial_l\Sigma\over
\partial K_A}>_J\, J_A- (-1)^{\varepsilon_A}
<{\partial_r {\cal O}\over \partial K_A}>_J\, J_A\nonumber\\&=&
- (-1)^{\varepsilon_A}{\partial_r <{\cal O}>_J\over \partial K_A}
J_A+{i\over \hbar}<{\cal O}>_J
{\partial_l W\over \partial K_A}J_A=
-(-1)^{\varepsilon_A}{\partial_r <{\cal O}>_J\over \partial K_A}J_A,
\end{eqnarray}
where we have used the Ward identity (\ref{ward2}) for $W$.
Using
\begin{equation}
\left. {\partial_r <{\cal O}>_J\over \partial K_A}\right|_J=
\left. {\partial_r <{\cal O}>_J\over \partial K_A}\right|_\Phi-
\left. {\partial_r <{\cal O}>_J\over \partial J_B}\right|_K
\left. {\partial_r J_B\over \partial K_A}\right|_\Phi,
\label{pr3}
\end{equation}
and some standard manipulation, one gets
\begin{equation}
0=(<{\cal O}>_J,\Gamma)-\left. {\partial_r <{\cal O}>_J
\over \partial J_B}\right|_K(J_B,\Gamma)=
(<{\cal O}>_J,\Gamma),
\end{equation}
which is the claimed result. We have used the fact that
\begin{equation}
(J_B,\Gamma)=0,
\label{gammajb}
\end{equation}
as one can prove
by differentiating (\ref{wardg}) with respect to $\Phi^B$.
Notice that a nontrivial formula like
(\ref{gammajb}) is once more consequence of the
nontrivial content of (\ref{wardg}). No similar formula can be derived
in classical or quantum mechanics.

With a similar argument, one can prove that if ${\cal O}$ is $\Omega$-exact,
then $<{\cal O}>_J$ is ${\rm ad}\, \Gamma$-exact.

We are thus allowed to define an observable
as an ${\rm ad}\,\Gamma_\infty$-closed
finite functional ${\cal O}$,
where $\Gamma_\infty$ denotes the finite
effective action (corrected at any loop
order in $\hbar$). We shall say more about this definition of observable
in the following.
It is the analogue of the classical concept of invariant of motion
or the quantum mechanical concept of observable.
Notice that ${\cal O}$ is defined up to ${\rm ad}\,\Gamma_\infty$-exact
terms. The nice property is that an ${\rm ad}\,\Gamma_\infty$-exact
functional $\Lambda$ is zero on shell
(i.e.\ at $J_A=0$ and $K_A=0$
or, equivalently, at $\Phi^A=0$ and $K_A=0$).
This is because if $\Lambda$ is ${\rm ad}\, \Gamma_\infty$-exact, then
there exists a ${\cal Q}$
such that $\Lambda=\left. {\partial_r {\cal Q}\over \partial K_A}\right|_JJ_A
(-1)^{\varepsilon_A}$.

\section{Removal of divergences in a generic gauge field theory}
\label{nonrenormalizable}

Let us now consider a generic nonrenormalizable gauge field theory. The
classical Lagrangian ${\cal L}_{class}$
will be supposed to be the most general gauge-invariant
one. The constants that
multiply the possible gauge-invariant terms of the classical Lagrangian
(infinite in number) will
be denoted by $\lambda$.

As we anticipated in the introduction,
we suppose that the zero-th order action $\Sigma_0$
is linear in the BRS
sources $K$,
\begin{equation}
\Sigma_0={\cal  L}_{class}(\phi)+s\Psi(\Phi)+K_As\Phi^A,
\label{azione}
\end{equation}
and moreover that the functional measure is BRS invariant,
i.e.\ $\Delta \Sigma_0=0$.

The algorithm that removes the divergences is derived by induction.
Quantities with a subscript $k$ refer to the theory in which the divergences
have been removed up to the $k^{th}$-loop order (for example $Z_k$,
$W_k$, $\Gamma_k$, $\Sigma_k$, \ldots).

The inductive hypothesis is the following. We assume that the
operator ${\cal L}_{n-1}$ that removes the $(n-1)^{th}$-order divergences
(when the lower order ones have already been removed) is a
transformation acting on $\lambda$ and $\{\Phi,K\}$ in
the following way
\begin{equation}
\matrix{
{\cal L}_{n-1}\lambda=\lambda+\delta_{n-1}\lambda,\cr
{\cal L}_{n-1}\{\Phi^A,K_A\}={\cal C}_{n-1}\{\Phi^A,K_A\},}
\end{equation}
where ${\cal C}_{n-1}$ denotes a local canonical transformation
(${\cal C}_{n-1}=1+{\cal O}(\hbar^{n-1})$), while
$\delta_{n-1}\lambda={\cal O}(\hbar^{n-1})$ denote
the $(n-1)^{th}$-loop order corrections of the constants $\lambda$.
The action of ${\cal L}_{n-1}$ on $\Sigma_{n-2}$ is
\begin{equation}
{\cal L}_{n-1}\Sigma_{n-2}(\Phi,K,\lambda)=
{\cal C}_{n-1}\Sigma_{n-2}(\Phi,K,\lambda+\delta_{n-1}\lambda).
\label{ln-1}
\end{equation}
The actions on $W_{n-2}$ and $\Gamma_{n-2}$ to
give $W_{n-1}$ and $\Gamma_{n-1}$, respectively,
are direct consequences of (\ref{ln-1}).

Let us define
\begin{equation}
{\cal R}_{n-1}={\cal L}_{n-1}\circ \cdots \circ {\cal L}_1=
{\cal L}_{n-1}\circ {\cal R}_{n-2}.
\end{equation}
${\cal R}_{n-1}$ acts on $\Sigma_0$, $W_0$ and $\Gamma_0$ and
gives $\Sigma_{n-1}$, $W_{n-1}$ and $\Gamma_{n-1}$.
It removes the divergences up to the $(n-1)^{th}$ loop order.
In general,
${\cal R}_{n-1}$ is not directly defined on $\lambda$ and $\{\Phi,K\}$.
It is a composition of redefinitions of $\lambda$
and local canonical transformations
on $\{\Phi,K\}$.
Notice that ${\cal C}_k$, ${\cal L}_k$ and ${\cal R}_k$ preserve locality
$\forall k$.
So, $\Sigma_k$ is local.

We are thus assuming that the divergences up to the $(n-1)^{th}$-loop
order have been corrected by a set of redefinitions of
the constants $\lambda$ and local canonical
transformations of fields and BRS sources. This guarantees that
$\Sigma_{n-1}$ satisfies the master equation, since $\Sigma_0$ does.
To show this, let us go back to the proof that a canonical transformation
preserves the master equation [see formul\ae\
(\ref{utility}), (\ref{utility2}) and (\ref{utility3})].
We have to improve it with
a redefinition of the constants $\lambda$.
This is extremely simple. Consider (\ref{utility}).
It was derived from (\ref{masterequation})
by changing names to fields and BRS sources
(from unprimed to primed objects). A similar
identity can be derived by changing names to fields, BRS sources and
the constants $\lambda$
(i.e.\ replacing $\lambda$ with $\lambda+\delta\lambda$).
None of the remaining steps of the proof gets
affected. With this simple improvement, all the properties
that hold for canonical transformations can be extended to
${\cal L}_{k}$-transformations, apart from those concerning
the derivatives with respect to $g$ (indeed,
the parameters $g$ can also enter
in the redefinitions of $\lambda$).
So, for example,
${\cal L}_k$-transformations preserve $\Omega$-closure
and $\Omega$-exactness, as well as
the Ward identity (\ref{wardg}) and so on.
Consequently, the same holds for any ${\cal R}_k$ and, in particular,
for ${\cal R}_{n-1}$. We are going to prove that the $n^{th}$-order
correction is of the same type. In other words, we have to define
suitable ${\cal L}_n$ and ${\cal R}_n$ operators.

When we speak
about {\sl redefinitions}
of the parameters $\lambda$, we are assuming that the classical
Lagrangian ${\cal L}_{class}$ is the most general one, i.e.\ that
it contains any possible $\lambda$.
Indeed, if we miss one, we cannot redefine it. So, when we want to consider a
classical Lagrangian with some missing terms, we have to set the
corresponding constants $\lambda$ to zero at the very end. The present
point is a key one: when the
nonrenormalizability is
only due to the gauge-fixing, ${\cal L}_{class}$ is not
the most general classical Lagrangian and one has to be sure that the missing
terms do not appear when applying the correction algorithm.

Now, we have a functional $\Gamma_{n-1}$ that is convergent up to the
$(n-1)^{th}$-loop order. Let us call $\Gamma^{(n)}_{div}$ its
$n^{th}$-order divergent part. $\Gamma^{(n)}_{div}$ is a
local functional. We want to show that we are able
to remove $\Gamma^{(n)}_{div}$ with a canonical transformation
${\cal C}_n=1+{\cal O}(\hbar^n)$
and a redefinition
$\lambda\rightarrow \lambda+\delta_n\lambda=\lambda+{\cal O}(\hbar^n)$
of the constants $\lambda$. This reproduces the inductive
hypothesis up to the $n^{th}$-loop order. As a consequence of the inductive
proof, there will exist a ${\cal R}$-transformation
(${\cal R}_{\infty}$, to be precise) that is able to remove all the divergences
of the theory.

Since $\Sigma_{n-1}$ satisfies the master equation (\ref{masterequation}),
then $\Gamma_{n-1}$ satisfies a Ward identity analogous to (\ref{wardg}),
namely
\begin{equation}
(\Gamma_{n-1},\Gamma_{n-1})=0.
\end{equation}
Taking the divergent part of the $n^{th}$-loop order of this expression,
we get
\begin{equation}
(\Gamma^{(n)}_{div},\Sigma_0)\equiv \sigma \Gamma^{(n)}_{div}=0.
\label{formu}
\end{equation}
Notice that $\sigma$ is nilpotent: $\sigma^2=0$, due to
$(\Sigma_0,\Sigma_0)=0$.
(\ref{formu}) is a very useful
characterization of $\Gamma^{(n)}_{div}$. As a matter
of fact, the general solution of this equation on local functionals
of zero ghost number
\cite{itzykson,justin,stelle,kluberg,joglekar} is
\begin{equation}
\Gamma^{(n)}_{div}={\cal G}^{(n)}(\phi)+(R^{(n)},\Sigma_0),
\label{divergences}
\end{equation}
for some local gauge-invariant functional ${\cal G}^{(n)}(\phi)$
depending only on the classical fields $\phi$ and a suitable local
functional $R^{(n)}(\Phi^A,K_A)$.
For Yang-Mills theories, this statement was conjectured in
ref.\ \cite{kluberg} and proved
in detail  in ref.\ \cite{joglekar}. The analogous statement
for higher derivative quantum gravity
was assumed without proof in ref.\ \cite{stelle}.
In this paper we assume
the same property without proof for our gauge field theory.
If formula (\ref{divergences}) were violated, for example if ${\cal G}^{(n)}$
depends on more fields than the classical ones, this could simply mean that
one has to enlarge the definition of ``classical fields''
and ``classical Lagrangian'' in order to include ${\cal G}^{(n)}$-type
functionals. Serious troubles, instead, can appear if ${\cal G}^{(n)}$
cannot be made independent of the BRS sources $K$: some arguments of our
derivations should be reconsidered. See also \cite{vandoren}.

${\cal G}^{(n)}(\phi)$ has the same form as ${\cal L}_{class}$,
of course,
since we are assuming that the classical Lagrangian
is the most general one.
So, ${\cal G}^{(n)}$ is responsible for the corrections
$\delta_n\lambda$ of the
constants $\lambda$. In other words, there exist suitable $\delta_n\lambda$
such that the corrected action $\Sigma_{n-1}(\Phi,K,
\lambda+\delta_n\lambda)$
gives a $\Gamma$-functional whose $n^{th}$-loop divergent part
is $\sigma$-exact: ${\cal G}^{(n)}$ is removed from
$\Gamma^{(n)}_{div}$ and only $(R^{(n)},\Sigma_0)$ remains.
This last piece can in fact be removed by a local canonical transformation
${\cal C}_n$
with a generating functional equal to
\begin{equation}
F^{(n)}(\Phi,K^\prime)=\Phi^AK_A^\prime+R^{(n)}(\Phi,K^\prime).
\label{fn}
\end{equation}
Notice that in the argument of $R^{(n)}$ the BRS sources are $K^\prime$
and not $K$.
In this way, it is easy to check that
\begin{equation}
\Sigma_n={\cal C}_n\Sigma_{n-1}(\Phi,K,\lambda+\delta_n\lambda)=
\tilde\Sigma_{n-1}+{i\hbar \over 2}\ln J_n=
\Sigma_{n-1}-\Gamma^{(n)}_{div}+{\cal O}(\hbar^{n+1}),
\label{4.6}
\end{equation}
where now the tilde on $\Sigma_{n-1}$
means not only that the fields and BRS sources
are substituted by their primed versions, but also that the constants
$\lambda$ are substituted by $\lambda+\delta_n \lambda$.
In deriving (\ref{4.6}),
we have used the fact that, since $R^{(n)}$ is of order $\hbar^n$,
(\ref{cantrasf}) give
\begin{equation}
\matrix{
\Phi^{\prime A}=\Phi^A+{\partial R^{(n)}\over \partial K_A}+
{\cal O}(\hbar^{n+1}),&
K^\prime_A=K_A-{\partial R^{(n)}\over \partial \Phi^A}+
{\cal O}(\hbar^{n+1}).}
\end{equation}
Moreover, $\ln \, J_n={\cal O}(\hbar ^n)$.
We conclude that $\Gamma_n$ is convergent up to the $n^{th}$-loop order.

The composition of the $n^{th}$-loop order correction ${\cal L}_n$
with
the operation ${\cal R}_{n-1}$ of removal of the lowest order divergences
defines an operator ${\cal R}_n={\cal L}_n\circ {\cal R}_{n-1}$,
that acts on the zero-th
order theory and removes all the divergences up to the $n^{th}$-loop order.
Moreover, ${\cal L}_n$ has the same structure as ${\cal L}_{n-1}$,
namely it is a redefinition of the constants $\lambda$
and a local canonical transformation ${\cal C}_n$.

According to the results of the previous section,
we have shown that there exists
an operator
${\cal R}_\infty$ that is able to remove all the divergences, while
preserving suitable extensions of
gauge-invariance, namely $\Omega_\infty$-invariance
on local functionals and ${\rm ad}\,\Gamma_\infty$-invariance on their
average values
(which are also generalizations of BRS-invariance).

Finally, notice that the canonical transformation (\ref{fn}) is not uniquely
fixed.
Any higher order correction can be introduced without affecting the
results. Moreover, due to (\ref{divergences}), $R^{(n)}$ is defined up to
$\sigma$-closed local functionals $T^{(n)}$, that can also be of
order $\hbar^n$.

\section{Predictivity with a nonrenormalizable gauge-fixing}
\label{predictivity}

The purpose of the present section is to improve the argument of the previous
one
in order to show that when a
renormalizable
theory is treated with a nonrenormalizable gauge-fixing,
predictivity is retained.
Now, the classical Lagrangian ${\cal L}_{class}$
is not the most general one, since it has to be
renormalizable.
The parameters $\lambda$ are finite in number.
The $\kappa$-dependence is entirely due to the gauge-fermion
$\Psi$. So, the previous argument can be adapted to the present case,
only if we are able to prove that
${\cal G}^{(n)}(\phi)$ is $\kappa$-independent,
so that it is only made of renormalizable terms, i.e.\
terms contained in the starting renormalizable classical Lagrangian
${\cal L}_{class}(\phi,\lambda)$.
Indeed, only in that case ${\cal G}^{(n)}$ can be absorbed with redefinitions
$\delta_n\lambda$ of the parameters $\lambda$ of ${\cal L}_{class}$.

Again, we proceed by induction.
Let us suppose that the algorithm works well
up to the $(n-1)^{th}$-loop order,
i.e.\ that $\forall k=1,\ldots n-1$, ${\cal L}_k$ is made by
$\kappa$-independent redefinitions $\delta_k\lambda$
of the parameters $\lambda$ and a
canonical transformation ${\cal C}_k$.
Formula (\ref{SIGMAPRIMOSULAMBDA}) for the change
of $\partial \Sigma\over\partial g$
under a canonical transformation is extendable to any
${\cal L}_k$-transformation, $k=1,\ldots n-1$, if we take $g=\kappa$, since
the correction of the parameters $\lambda$ is $\kappa$-independent
by inductive hypothesis. So, if ${\partial \Sigma_{k-1}\over \partial \kappa}$
is $\Omega_{k-1}$-exact, then ${\partial \Sigma_k\over \partial \kappa}$
is $\Omega_k$-exact. A similar property extends to ${\cal R}_{n-1}$:
if ${\partial \Sigma_0\over \partial \kappa}$
is $\Omega_0$-exact, then ${\partial \Sigma_{n-1}\over \partial \kappa}$
is $\Omega_{n-1}$-exact.

Notice that in a renormalizable gauge-field theory that is
gauge-fixed with a nonrenormalizable gauge-fixing,
${\partial \Sigma_0\over\partial \kappa}$ is BRS-exact. As a matter of fact,
\begin{equation}
\Sigma_0={\cal L}_{class}(\phi,\lambda)+s\Psi(\Phi,\kappa)+K_As\Phi^A,
\end{equation}
$\Psi$ being the gauge-fermion, and so,
\begin{equation}
{\partial \Sigma_0\over\partial \kappa}=s{\partial \Psi\over \partial \kappa}.
\end{equation}
Since $\Psi$ is independent
of the BRS sources $K$, we can also write
\begin{equation}
{\partial \Sigma_0\over\partial \kappa}=\left(
{\partial \Psi\over \partial \kappa},\Sigma_0\right)=\Omega_0
{\partial \Psi\over \partial \kappa},
\label{5.3}
\end{equation}
$\Omega_0$ denoting the zeroth order $\Omega$-operator.
Since ${\partial \Sigma_0\over
\partial \kappa}$ is local and $\Omega_0$-exact, then
(\ref{SIGMAPRIMOSULAMBDA})
and the above remarks
assure that
${\partial \Sigma_{n-1}\over \partial
\kappa}$ is local and $\Omega_{n-1}$-exact. Let $\chi_{n-1}$ be such that
\begin{equation}
{\partial \Sigma_{n-1}\over \partial \kappa}=\Omega_{n-1}\chi_{n-1}.
\label{sigmasueexact}
\end{equation}
Since $\chi_{n-1}={\cal R}_{n-1}{\partial\Psi\over\partial \kappa}$,
we see that $\chi_{n-1}$ is a local functional.
(\ref{sigmasueexact}) is sufficient to prove that
${\partial \Gamma_{n-1}\over \partial \kappa}$ is ${\rm
ad}\,\Gamma_{n-1}$-exact,
which can be obtained following the same steps of the proof that
if ${\cal O}$ is $\Omega$-closed, then $<{\cal O}>_J$
is ${\rm ad}\,\Gamma$-closed.

Indeed, let us start from
\begin{equation}
{\partial W_{n-1}\over \partial \kappa}=<{\partial\Sigma_{n-1}\over \partial
\kappa}>_J=<\Omega_{n-1}\chi_{n-1}>_J.
\label{dewsudee}
\end{equation}
By performing the change of variables
\begin{equation}
\delta\Phi^A={\partial _l\Sigma_{n-1}\over \partial K_A}\Lambda,
\end{equation}
in the functional integral $<\chi_{n-1}>_J$,
one finds a formula analogous to (\ref{pr1}), with $\chi_{n-1}$
in replacement of ${\cal O}$.
Moreover, with an integration by parts, one can prove the
analogue of (\ref{pr2}).
It is thus possible to write (\ref{dewsudee}) in the form
\begin{equation}
{\partial W_{n-1}\over \partial \kappa}=
{\partial_r <\chi_{n-1}>_J\over \partial K_A}J_A(-1)^{\varepsilon_A}.
\end{equation}
Now, the definition of $\Gamma_{n-1}$ [see (\ref{gammadefinition})],
permits to write
\begin{equation}
\left. {\partial\Gamma_{n-1}\over \partial \kappa}\right|_{\Phi,K}=
\left. {\partial W_{n-1}\over \partial \kappa}\right|_{J,K}=
\left. {\partial_r <\chi_{n-1}>_J\over \partial K_A}
\right|_JJ_A(-1)^{\varepsilon_A}.
\end{equation}
Using the analogue of (\ref{pr3})
and (\ref{gammajb}), one gets
\begin{equation}
{\partial \Gamma_{n-1}\over \partial \kappa}=(
<\chi_{n-1}>_J,\Gamma_{n-1}).
\label{gammasueexact}
\end{equation}
We conclude that ${\partial \Gamma_{n-1}
\over \partial \kappa}$ is ${\rm ad}\,\Gamma_{n-1}$-exact.

Let us call $S_{n-1}=<\chi_{n-1}>_J$.
$S_{n-1}$ is the average value (at nonzero sources $J$)
of a local functional.
Notice that $S_{n-1}$ is determined up to additions of
${\rm ad}\,\Gamma_{n-1}$-closed functionals.
We introduce the additional inductive hypothesis that
$S_{n-1}$ is finite up to the $(n-1)^{th}$-loop order.
Of course, this hypothesis is satisfied at lowest order:
$S_0=<\chi_0>_J=<{\partial\Psi\over \partial \kappa}>_J=
{\rm finite}\, +{\cal O}(\hbar)$.
We shall have to prove that the additional hypothesis
is reproduced to the $n^{th}$ loop order, i.e.\ that
$S_n=<\chi_n>_J$ can be chosen finite up to order $\hbar^n$.

Let $S^{(n)}_{div}$ denote the $n^{th}$-loop order divergent part
of $S_{n-1}$. $S_{div}^{(n)}$ is local, since
we are assuming that all the subdivergences have been removed.
Let us focus on the $n^{th}$-loop order divergent part of equation
(\ref{gammasueexact}), namely
\begin{equation}
{\partial \Gamma^{(n)}_{div}\over \partial \kappa}=
(\chi_0,\Gamma^{(n)}_{div})+(S^{(n)}_{div},\Sigma_0).
\end{equation}
On the other hand, (\ref{divergences}) gives
\begin{equation}
{\partial \Gamma^{(n)}_{div}\over \partial \kappa}=
{\partial {\cal G}^{(n)}\over \partial \kappa}+
\left(R^{(n)},{\partial \Sigma_0\over \partial \kappa}\right)+
\left({\partial R^{(n)}\over \partial \kappa},\Sigma_0\right),
\end{equation}
so that we can write
\begin{equation}
{\partial {\cal G}^{(n)}\over \partial \kappa}=
(\chi_0,\Gamma^{(n)}_{div})-
\left(R^{(n)},{\partial \Sigma_0\over \partial
\kappa}\right)+(\ldots,\Sigma_0).
\end{equation}
Using (\ref{5.3}), we get
\begin{equation}
{\partial {\cal G}^{(n)}\over \partial \kappa}=
(\chi_0,{\cal G}^{(n)})+(\ldots,\Sigma_0).
\label{gnsue}
\end{equation}
Now, since $\chi_0$ and ${\cal G}^{(n)}(\phi)$ do not depend on the BRS
sources $K$, we have
\begin{equation}
(\chi_0,{\cal G}^{(n)})=0,
\end{equation}
and we can conclude
that $\partial {\cal G}^{(n)}\over \partial \kappa$ is $\sigma$-exact
and that it is equal to the action of $\sigma$ on
a local functional.
This implies that it can only be zero\footnotemark
\footnotetext{See \cite{joglekar}. As a matter of fact, one can always
assume that, in formula (\ref{divergences}),
${\cal G}^{(n)}$ does not contain any
gauge-invariant term of the form $({\sl local \, functional},
\Sigma_0)$. This simply amounts to a redefinition of $R^{(n)}$.
If this is the case, then ${\partial {\cal G}^{(n)}\over \partial \kappa}$ is
also a sum of gauge-invariant terms, none of which
can be written as $({\sl local \, functional},
\Sigma_0)$.}.
This proves that the redefinitions $\delta_n\lambda$
of the constants $\lambda$
are
$\kappa$-independent.

The $\kappa$-independence of $\delta_n\lambda$ generalizes the well-known
gauge-independence of the coupling constant renormalization. See also
section \ref{examples} for a comment about this fact within Yang-Mills theory
treated in the usual way.

As a consequence of ${\partial \delta_n\lambda\over
\partial \kappa}=0$, we can write
\begin{equation}
{\partial\Sigma_n\over \partial \kappa}=
\Omega_n\chi_n,
\label{uno}
\end{equation}
and
\begin{equation}
{\partial \Gamma_n\over \partial \kappa}=(S_n,\Gamma_n),
\label{due}
\end{equation}
being $S_n=<\chi_n>_J$. Formula (\ref{SIGMAPRIMOSULAMBDA}) gives
\begin{equation}
\chi_n=\tilde\chi_{n-1}-{\partial F^{(n)}\over \partial \kappa}
\label{tre}
\end{equation}
(the tilde, as usual now, means that the parameters $\lambda$ have to be
substituted with the new ones and
the fields and BRS sources have to be substituted with the
canonically transformed ones). Moreover,
$S_n=<\chi_n>_J=S_{n-1}+{\cal O}(\hbar^n)$. Consequently,
$S_n$ is surely finite up to order $\hbar^{n-1}$. In order to fully
reproduce the inductive hypothesis, we have to prove that $S_n$ can
be chosen finite up to order $\hbar^n$.

Let ${\cal S}^{(n)}_{div}$ denote the $n^{th}$ loop order divergent
part of $S_n$. ${\cal S}^{(n)}_{div}$ is local, since the subdivergences
vanish by inductive assumption.
Taking the $n^{th}$ loop order divergent part of (\ref{due}),
one has
\begin{equation}
\sigma {\cal S}^{(n)}_{div}=0.
\label{quattro}
\end{equation}

We shall give two different arguments for removing ${\cal S}^{(n)}_{div}$.
The first method is based on the fact that $S_n$ is defined up to
${\rm ad}\, \Gamma_n$-closed functionals, that are averages of
$\Omega_n$-closed local functionals, while the second method is based
on the fact that the generating functional $F^{(n)}$ of (\ref{fn})
is defined up to $\sigma$-closed functionals $T^{(n)}={\cal O}(\hbar^n)$.

Let us begin with the first method. It is fundamental to
be able to extend ${\cal S}^{(n)}_{div}$ to an $\Omega_0$-closed local
functional ${\cal S}^{(n)\prime}_{div}$ by adding higher
order terms. Were ${\cal S}^{(n)}_{div}$ of ghost number zero,
this would be very easy: ${\cal S}^{(n)}_{div}$ would be of the form
\begin{equation}
{\cal S}^{(n)}_{div}=f(\phi)+(h,\Sigma_0)=f+\sigma h,
\end{equation}
where $f$ is a gauge-invariant local functional of the classical fields $\phi$
and $h$ is local.
Of course, $f$ and $h$ are of order $\hbar^n$, as ${\cal S}^{(n)}_{div}$.
Then, the $\Omega_0$-closed extension of ${\cal S}^{(n)}_{div}$
is
\begin{equation}
{\cal S}^{(n)\prime}_{div}\equiv f+\Omega_0 h={\cal S}^{(n)}_{div}+
{\cal O}(\hbar^{n+1}).
\end{equation}
However, ${\cal S}^{(n)}_{div}$ has ghost number $-1$ and it is
not simple to estabilish the cohomology content of $\sigma$ on
ghost number $-1$ local functionals. Nevertheless, due to
$\Omega_0=\sigma-i\hbar \Delta$, one can notice that at least in
a dimensional regularization framework,
${\cal S}^{(n)\prime}_{div}={\cal S}^{(n)}_{div}$
is trivially $\Omega_0$-closed.
Then, we know from section \ref{observables} and the remarks
of section \ref{nonrenormalizable},
that the operator ${\cal R}_{n}$ permits to find a local
$\Omega_{n}$-closed extension ${\cal R}_{n}{\cal S}^{(n)\prime}_{div}
={\cal S}^{(n)}_{div}+{\cal O}(\hbar^{n+1})$
of ${\cal S}^{(n)\prime}_{div}$.
Finally, we know from section \ref{observables} that
\begin{equation}
{\cal S}_{n}\equiv<{\cal R}_{n}{\cal S}^{(n)\prime}_{div}>_J
\end{equation}
is ${\rm ad}\,\Gamma_{n}$-closed. Moreover, ${\cal S}_{n}=
{\cal S}^{(n)}_{div}+{\cal O}(\hbar^{n+1})$.
Consequently, ${\cal S}_n$ can be safely subtracted from $S_n$:
this cancels the divergent part ${\cal S}^{(n)}_{div}$ and preserves
(\ref{due}). The subtraction of ${\cal S}_n$ from $S_n$
corresponds to a subtraction of
${\cal R}_n{\cal S}^{(n)\prime}_{div}$ from $\chi_n$: (\ref{uno})
is also preserved.

The second method does not require any restriction on the regularization
technique. We know that $F^{(n)}$ of formula (\ref{fn}) is defined up
to $\sigma$-colsed functionals $T^{(n)}$. The addition of
such $T^{(n)}$'s to $F^{(n)}$ only changes
$\Sigma_{n}$ and $\Gamma_n$ to order
$\hbar^{n+1}$ and neither affects (\ref{uno}) nor
(\ref{due}).  However, due to (\ref{tre}), it affects $\chi_n$ and also
$S_n=<\chi_n>_J$. Moreover, $T^{(n)}$ has ghost
number $-1$: it is thus a good candidate for our purposes.
Let us choose $T^{(n)}$ to be ${\cal O}(\hbar^n)$ and divergent.
(\ref{tre}) assures that when introducing a $T^{(n)}$ in $F^{(n)}$,
$\chi_n$ goes into $\chi_n-{\partial T^{(n)}\over \partial \kappa}-
(T^{(n)},\chi_0)+{\cal O}(\hbar^{n+1})$ and that
${\cal S}^{(n)}_{div}$ changes according to
\begin{equation}
{\cal S}^{(n)}_{div}\rightarrow {\cal S}^{(n)}_{div}-
{\partial T^{(n)}\over \partial \kappa}-(T^{(n)},\chi_0).
\label{cinque}
\end{equation}
$\sigma T^{(n)}=0$ also implies
\begin{equation}
0={\partial (\sigma T^{(n)})\over \partial \kappa}=\sigma
\left({\partial T^{(n)}\over \partial \kappa}+(T^{(n)},\chi_0)\right),
\end{equation}
as it must be, due to (\ref{quattro}). One would like to find a $T^{(n)}$
such that the right hand side of (\ref{cinque}) is zero. This
can be done perturbatively in $\kappa$. Let us start with
$T^{(n)}=\kappa {\cal S}^{(n)}_{div}$:
$\sigma (\kappa {\cal S}^{(n)}_{div})=0$. Then
\begin{equation}
{\cal S}^{(n)}_{div}\rightarrow
-\kappa {\partial {\cal S}^{(n)}_{div}\over \partial \kappa}-\kappa
({\cal S}^{(n)}_{div},\chi_0)\equiv \kappa {\cal T}^{(1)}.
\end{equation}
${\cal S}^{(n)}_{div}$ has not been set to zero, however its power expansion
in $\kappa$ starts now with order one. Clearly, $\sigma {\cal T}^{(1)}=0$,
due to (\ref{quattro}). Let us now add ${\kappa^2\over 2}{\cal T}^{(1)}$
to $T^{(n)}$. We have
\begin{equation}
{\cal S}^{(n)}_{div}\rightarrow
-{\kappa^2\over 2}\left(
{\partial {\cal T}^{(1)}\over \partial \kappa}+({\cal T}^{(1)},\chi_0)\right)
\equiv \kappa^2 {\cal T}^{(2)}.
\end{equation}
Again, $\sigma {\cal T}^{(2)}=0$. Moreover,
${\cal S}^{(n)}_{div}$ has become quadratic in $\kappa$. Then we can add
${\kappa^3\over 3}{\cal T}^{(2)}$ to $T^{(n)}$ and go on:
${\cal S}^{(n)}_{div}$ can be made of arbitrarily high order
in $\kappa$ (always of order $n$ in $\hbar$)
and so we can conclude that it can be made to vanish, as desired.

A key remark, now, is the following. The proof that we have made
can be extended to any nonrenormalizable gauge field theory, to
show that if some parameter $g$ (not necessary of
negative dimension in mass units) only appears in the gauge-fermion
$\Psi$, but neither in the classical Lagrangian
${\cal L}_{class}(\phi,\lambda)$,
nor in the BRS variations $s\Phi^A$ of the fields $\Phi^A$, then
the order by order redefinitions of the parameters $\lambda$
of ${\cal L}_{class}$ (now infinitely many) is $g$-independent
and the derivative of $\Gamma_\infty$
with respect to $g$ is ${\rm ad}\,\Gamma_\infty$-exact.

This is useful for the definition of the observables of a
{\sl renormalizable} field theory. Let $\{{\cal O}_i(\phi)\}$ be a
basis of local gauge-invariant operators, constructed only
with the classical fields $\phi$. Let us substitute the classical Lagrangian
${\cal L}_{class}(\phi,\lambda)$ with
\begin{equation}
{\cal L}^{(\beta)}_{class}(\phi,\lambda,\beta)
\equiv {\cal L}_{class}(\phi,\lambda)+\beta_i{\cal O}_i(\phi).
\label{5.21}
\end{equation}
The $\beta_i$ can be point-dependent. For simplicity, such a dependence
will be undestrood.
The Lagrangian (\ref{5.21})
is the same as a nonrenormalizable Lagrangian, so that the correction
algorithm permits to define a finite functional
\begin{equation}
\Gamma_\infty^{(\beta)}(\Phi,K,\lambda,\kappa,\beta)
\end{equation}
such that its derivative with respect to $\kappa$ is
${\rm ad}\,\Gamma_\infty$-exact and the order by order
redefinitions
of $\lambda$ and $\beta$ are $\kappa$-independent.
The only
(fundamental) difference with
a true nonrenormalizable
field theory is that $\beta_i$ are not true coupling constants,
but artificial parameters that permit to define  the
quantum extensions ${\cal O}_i^{(q)}$ of
the observables ${\cal O}_i$, that are
\begin{equation}
{\cal O}_i^{(q)}(\Phi,K)=
\left. {\partial \Gamma^{(\beta)}_\infty\over\partial
\beta_i}\right|_{\beta=0}.
\end{equation}
The quantum generalization of the amplitude
$<{\cal O}_{i_1}\cdots {\cal O}_{i_n}>$ is given by
\begin{equation}
{\cal A}_{i_1\cdots i_n}\equiv \left. {\partial^n\Gamma_\infty^{(\beta)}
\over \partial \beta_{i_1}\cdots \partial \beta_{i_n}}\right|_{\beta=0,\,
\Phi=0,\,K=0}.
\end{equation}
We know that, on shell,
\begin{equation}
{\partial \Gamma_\infty^{(\beta)}\over \partial \kappa}=0
\end{equation}
and so we conclude that the on shell amplitudes
are $\kappa$-independent, namely
\begin{equation}
{\partial {\cal A}_{i_1\cdots i_n}\over \partial \kappa}=0.
\end{equation}
Differentiating $(\Gamma_\infty^{(\beta)},\Gamma_\infty^{(\beta)})=0$
with respect to $\beta_i$ and setting $\beta=0$,
we get
\begin{equation}
{\rm ad}\, \Gamma_\infty \, {\cal O}_i^{(q)}
=({\cal O}_i^{(q)},\Gamma_\infty)=0,
\end{equation}
where $\Gamma_\infty=\Gamma_\infty^{(\beta)}|_{\beta=0}$.
In other words, ${\cal O}_i^{(q)}$ are quantum
observables in the sense
defined in section \ref{observables}.
Moreover,
\begin{equation}
{\cal O}_i^{(q)}=
\left.\left(\left.{\partial \Gamma_\infty^{(\beta)}\over \partial \beta_i}
\right|_{\Phi,K}\right)\right|_{\beta=0}=
\left.\left(\left.{\partial W_\infty^{(\beta)}\over \partial \beta_i}
\right|_{J,K}\right)\right|_{\beta=0}=
\left. <{\partial\Sigma_\infty^{(\beta)}\over \partial \beta_i}>_J
\right|_{\beta=0}.
\end{equation}
The differentiation of the master equation for
$\Sigma_\infty^{(\beta)}$ with respect to $\beta_i$ implies
\begin{equation}
\Omega_\infty \left. {\partial\Sigma_\infty^{(\beta)}\over \partial \beta_i}
\right|_{\beta=0}=0,
\end{equation}
so that the quantum observables ${\cal O}_i^{(q)}$ are average values of
$\Omega_\infty$-closed local functionals, again in agreement
with the definitions developed in section \ref{observables}.

Concluding, when the nonrenormalizability is only due
to the gauge-fixing,
the physical amplitudes depend on finitely many parameters
and so a finite number of measurements is necessary to
determine uniquely the theory.

Again, the previous conclusions can also be extended to
a true nonrenormalizable field
theory: the physical amplitudes can depend on infinitely many
parameters, nevertheless they cannot depend on the
parameters that were introduced only through the gauge-fixing,
i.e.\ the independence from the gauge-fermion $\Psi$ survives the subtraction
algorithm.

\section{Examples}
\label{examples}

In this section we examine some examples of renormalizable
field theories that are treated with a nonrenormalizable gauge-fixing.
The first example is pure Q.E.D. (to further simplify things,
we consider the case of space-time dimension two):
it is a free theory and so there
is no nontrivial physical amplitude.
Let us regularize with the dimensional technique.
We choose the gauge-fixing
\begin{equation}
\partial_\mu A_\mu+\kappa(\partial _\mu A_\nu)(\partial_\mu A_\nu)=0.
\end{equation}
This is a continuous deformation of the usual gauge-fixing
$\partial_\mu A_\mu=0$,
so that we expect the same physical results.
The BRS algebra is
\begin{equation}
\matrix{sA_\mu=\partial_\mu c,&sc=0,\cr s\bar c=b,&sb=0,}
\end{equation}
with obvious notation.
We choose the following gauge fermion $\Psi$
\begin{equation}
\Psi={1\over 2}\bar c\,(b+2\partial\cdot A+2\kappa(\partial_\mu A_\nu)^2),
\end{equation}
so that the BRS action is
\begin{equation}
\Sigma_0=-{1\over 4}F_{\mu\nu}^2+s\Psi
+K_AsA+K_c sc+K_{\bar c}s \bar c+K_bsb.
\end{equation}
Notice that the Lagrange multiplier $b$ is
not integrated away. $A_\mu$ and $b$ must be treated as a whole
to define a nonsingular propagator. One then checks that there
is propagation from $A_\mu$ to $A_\nu$ and from
$A_\mu$ to $b$ (and {\sl viceversa}), but no
propagation from $b$ to $b$.
There is no dependence on $K_c$ and $K_b$,
since $sc=sb=0$. Moreover, there is no radiative correction to
the other BRS transformations, since they are linear (correspondingly
$K_AsA+K_{\bar c}s$ is quadratic). This implies that the linear dependence
of the action on the BRS sources is radiatively preserved\footnotemark
\footnotetext{Strictly speaking, one should say that the condition of linearity
in the BRS sources {\sl can} be radiatively preserved, i.e.\ that this
condition is compatible with the subtraction algorithm. Indeed,
if this restriction is not specified,
one could introduce arbitrary finite nonlinear terms.}.
Let us rewrite $\Sigma_0$ in a more explicit form, namely
\begin{eqnarray}
\Sigma_0&=&{1\over 2}A_\mu\Box A_\mu+{1\over 2}(\partial \cdot A)^2
+{1\over 2}b^2+b(\partial\cdot A
+\kappa(\partial_\mu A_\nu)^2)\nonumber\\&
-&\bar c (\Box c + 2\kappa \partial _\mu\partial_\nu c \, \partial_\mu A_\nu)
+K_A^\mu\partial_\mu c+ K_{\bar c}b.
\end{eqnarray}

Let us consider the one loop amplitudes with external photonic legs.
They are the sum of two diagrams: in one of them $A_\mu$ and $b$
circulate in the loop, while in the other one,
$\bar c$ and $c$ circulate. It is simple to check that these
two diagrams exactly cancel, so that the amplitude is identically zero,
independently of the number of external photonic legs.

Next, consider the one loop diagram with two external $b$-legs (of
momentum $p$ and $-p$):
there is a divergence of the kind
\begin{equation}
{1\over \varepsilon}\kappa^2 p^2.
\end{equation}
This means that it is necessary to introduce a counterterm of the form
\begin{equation}
{1\over 2}\kappa^2 b\Box b.
\end{equation}
We conclude that the Lagrange multiplier becomes ``propagating''.
Counterterms of the form $\kappa^3b^2\Box b$ and $\kappa^3b(\partial_\mu b)^2$
are also required, so that the corrected action is not even quadratic
in $b$.
That is why, for the simplicity of the computation,
it is preferable to avoid integrating $b$ away.

Instead, starting from a Lagrangian $\Sigma_0$ in which
$b$ has been integrated away, namely
\begin{eqnarray}
\Sigma_0&=&{1\over 2}A_\mu\Box A_\mu-\kappa\partial \cdot A
(\partial_\mu A_\nu)^2-
{1\over 2}\kappa^2((\partial_\mu A_\nu)^2)^2\nonumber\\&
-&\bar c (\Box c + 2\kappa \partial _\mu\partial_\nu c \, \partial_\mu A_\nu)
+K_A^\mu\partial_\mu c+ K_{\bar c}b,
\end{eqnarray}
it is no more true that the one loop
amplitudes
with only photonic external legs
\begin{equation}
<A_{\mu_1}(p_1)\cdots A_{\mu_n}(p_n)>\Big|_{\rm one \, loop}
\end{equation}
are zero. Nevertheless, it is true that the physical projections
\begin{equation}
\left( \delta_{\mu_1\nu_1}-{p_{1\mu_1}p_{1\nu_1}\over p_1^2}
\right)\cdots \left( \delta_{\mu_n\nu_n}-{p_{n\mu_n}p_{n\nu_n}
\over kp_n^2}\right)
<A_{\nu_1}(p_1)\cdots A_{\nu_n}(p_n)>\Big|_{\rm one \, loop}
\end{equation}
vanish. Check, for example, the case $n=2$.

Let us now couple fermions to the electromagnetic field, by
adding to $\Sigma_0$ the terms
\begin{equation}
\bar \psi (\partial\!\!\!\slash+ A\!\!\!\slash)\psi+
\bar Ds\psi+s\bar \psi D.
\end{equation}
The BRS algebra is
\begin{equation}
\matrix{s\psi=-c\psi,&s\bar\psi=c\bar\psi.}
\end{equation}
The nonlinearity of these transformations gives rise to nontrivial
radiative corrections to them
together with the lost of linearity in the BRS sources.
As a matter of fact, although it is simple to check that there is
no divergent one loop diagram with both
$\bar D$- and $D$-external legs,
nevertheless two loop divergent diagrams with both
$\bar D$- and $D$-external legs do exist.

Analogous considerations apply for non-abelian Yang-Mills theory.

Let us now describe where, in our description, the usual coupling
constant and wave function renormalization come from, when
a renormalizable gauge-field theory is treated with
one of the usual renormalizable gauge-fixings.
For simplicity, we consider pure non-abelian Yang-Mills theory.
In the usual approach, three independent renormalization constants $Z$
are needed: one for the coupling constant $g$, one for the wave-function
renormalization
of the vector $A_\mu$ and one for the
ghosts-antighost wave-function renormalization.
On the other hand, in our approach the removal of divergences
is performed by  a redefinition of the parameters that multiply
the gauge-invariant terms of the starting classical Lagrangian
(in the present case there is only one such term, namely $F^a_{\mu\nu}
F_{\mu\nu}^a$;
let us call $\lambda$ the constant in front of it) and
a canonical transformation of fields and BRS sources.
If the gauge-fixing is renormalizable, the canonical transformation
is linear,
i.e.\ the fields $\Phi^A$ and the BRS sources $K_A$
are simply multiplied by some constants, which we call
$Z_{\Phi^A}$ and $Z_{K_A}$, respectively.
The requirement of preservation of antibrackets implies
$Z_{\Phi^A}Z_{K_A}=1$ $\forall A$. This reduces the number
of independent $Z$-factors for fields and BRS sources to a half,
namely $4$ (those
for $A_\mu$, $c$, $\bar c$ and $b$, for instance).
However, since $K_{\bar c}b$ and $b\partial\cdot A$ are not radiatively
corrected, we have $Z_b=Z_{\bar c}={1\over Z_{A_\mu}}$.
Thus, we remain with three independent $Z$-factors: those
for $A_\mu$ and $c$ and that for $\lambda$. The correct counting is
thus retrieved and one can also check that these three $Z$-factors
are indeed sufficient to produce the usual coupling constant renormalization
and wave function renormalizations.
Notice that
there is no redefinition of the coupling constant $g$ in our description;
indeed the usual redefinition of $g$ is recovered
from the redefinitions of $\lambda$ and $A_\mu$.
Precisely, the usual renormalization factor $Z_g$ for the gauge-coupling
constant $g$ results to be equal to $Z_\lambda^{-1/2}$. Thus, the
gauge-independence of $Z_\lambda$, proved in section \ref{predictivity},
is nothing but the familiar gauge-independence of $Z_g$.

Let us now consider the topological $\sigma$-model
formulated in \cite{anselmifresigma}. It is an irreducible
gauge-field theory and
its BRS algebra [formula (13) of \cite{anselmifresigma}] can
be written, after a natural redefinition of the Lagrange multiplier
$b_\mu^i$, in the form
\begin{equation}
\matrix{
sq^i=\xi^i,&s\xi^i=0,\cr s\zeta_\mu^i=b_\mu^i,&s b_\mu^i=0.}
\end{equation}
The simplicity of the BRS algebra assures that linearity in the BRS
sources is preserved and that there is no radiative correction
to the BRS transformations. Thus, the observables that were
listed in \cite{anselmifresigma} are directly promoted to
quantum observables. The expression ${\cal G}^{(n)}$ appearing
in formula (\ref{divergences}) is zero,
since any gauge-invariant functional (i.e.\ a topological invariant)
is perturbatively trivial. This implies that there is no redefinition
of the parameters $\lambda$.
The expression $R^{(n)}$ appearing
in the same formula (\ref{divergences})
is independent of the BRS sources. This
implies that the canonical transformation that absorbs the
divergent terms leaves the fields invariant and only changes the
BRS sources according to
\begin{equation}
\delta K_A=-{\partial R^{(n)}\over \partial \Phi^A}.
\end{equation}
Formula (\ref{antifields}) shows that this is precisely a redefinion of the
gauge-fermion $\Psi$:
\begin{equation}
\Psi_{n-1}\rightarrow \Psi_n=\Psi_{n-1}-R^{(n)}.
\end{equation}
We conclude that the removal of divergences
simply reduces to a redefinition of the gauge-fermion and thus
has no physical consequence.

The nonrenormalizability of topological fields theories coming
from the twist of some N=2 nonrenormalizable quantum field theory is thus
turned into a positive feature: it shows that a suitable
subset of the physical amplitudes of
a nonrenormalizable N=2 quantum field theory is in any case predictive and
physically well-defined.

\section{Conclusions}
\label{conclusions}

Apart from eventual applications, the investigation about
the removal of divergences in a nonrenormalizable
gauge-field theory turns out to be enlightening rather than
extravagant. The usual theorem of renormalizability of Yang-Mills
theories is not, strictly speaking, a {\sl renormalizability} theorem,
since our improved version works for any (eventually
nonrenormalizable) gauge-field theory.
Rather, it is a theorem on the compatibility (up to BRS anomalies)
of gauge-invariance
with the subtraction algorithm, a fact that is
fundamental for unitarity. When combining that theorem
with power counting, one can determine the renormalizability or
nonrenormalizability of the theory. This is not the whole story about
predictivity. As a matter of fact, when the counterterms are infinitely
many, one has to determine how many of them are nontrivial
(i.e.\ non ``BRS exact''): if the nontrivial counterterms are
infinitely many, then the theory is not predictive.
If the nontrivial counterterms are finitely many, then the theory is
predictive. We have shown in detail that when nonrenormalizability
is only due to the gauge-fixing, predictivity is preserved,
a fact that is naturally expected.
One is lead to wonder whether there are more general cases
of predictive nonrenormalizability. This could require a revision
of our idea of physically acceptable field theories.

Finally, the simple description of the removal of divergences as a
redefinition of some parameters
together with a canonical transformation
of fields and BRS sources opens the possibility that in some simple
models such a set of redefinitions, or, in other words,
the identification of the {\sl correct}
variables and the correct parameters, is derivable from first
principles with a synthetic argument, i.e.\
without any analytic computation.

\vspace{24pt}
\begin{center}
{\bf Acknowledgements}
\end{center}

\vspace{12pt}
I would like to thank E.\ Witten and R.\ Stora for useful discussions.

\vspace{24pt}
\begin{center}
{\bf Appendix: Proof of formula (36)}
\end{center}
\vspace{12pt}

We want to prove formula (\ref{SIGMAPRIMOSULAMBDA}).
We find it convenient to use the notation of ref.\ \cite{diaz}
and some of the formul\ae\ proven there.
Let us define
\begin{equation}
\matrix{
{M^A}_B={\partial_l\partial F\over \partial K^\prime_A\partial\Phi^B},&
{N_B}^A={\partial_l\partial F\over \partial\Phi^B\partial K^\prime_A},\cr
F^{AB}={\partial_l\partial F\over \partial K^\prime_B\partial K^\prime_A},&
F_{AB}={\partial_l\partial F\over \partial\Phi^A\partial \Phi^B}.}
\end{equation}
The statistics are as follows
\begin{equation}
\matrix{
\varepsilon({M^A}_B)=\varepsilon({N_B}^A)=\varepsilon_A+\varepsilon_B,
&\varepsilon(F^{AB})=\varepsilon(F_{AB})=\varepsilon_A+\varepsilon_B+1.}
\end{equation}
The following rules for transposition hold:
\begin{equation}
\matrix{
{N_A}^B={M^B}_A (-1)^{\varepsilon_A (\varepsilon_B+1)},&
{(N^{-1})_A}^B={(M^{-1})^B}_A (-1)^{\varepsilon_A (\varepsilon_B+1)}.}
\end{equation}
The differentials of $\Phi^\prime$ and $K$ are
\begin{eqnarray}
d\Phi^{\prime A}&=&{M^A}_Bd\Phi^B+F^{AB}d K^\prime _B+
{\partial \Phi^{\prime A}\over \partial g}dg,\nonumber\\
dK_A&=&d\Phi^BF_{BA}+d K^\prime _B{M^B}_A+
{\partial K_A\over \partial  g}d g .
\label{differentials}
\end{eqnarray}
As a convention, when there cannot be any misunderstanding,
we do not specify the variables
that are taken to be constant in a partial derivative. It is
understood that the differentiated
function is considered as a function of its natural variables:
$F$, $\Phi^\prime$ and $K$ are functions of
$\{\Phi,K^\prime, g \}$ [see (\ref{cantrasf})], $\Sigma$ and
$\Sigma^\prime$ are functions of $\{\Phi,K, g \}$,
$\tilde\Sigma$ is a function of $\{\Phi^\prime,K^\prime, g \}$,
and so on.

Let us differentiate (\ref{sigmaprimo})
with respect to $g$ at constant $\Phi$ and $K$.
\begin{equation}
{\partial \Sigma^\prime\over \partial g}=
{\partial \tilde\Sigma\over \partial g}
(\Phi^\prime(\Phi,K),K^\prime(\Phi,K))
+\left. {\partial \Phi^{\prime A}\over
\partial g}\right|_{\Phi,K}
{\partial_l \tilde \Sigma\over \partial\Phi^{\prime A}}+
\left. {\partial K^\prime_A\over
\partial g}\right|_{\Phi,K}{\partial_l \tilde \Sigma
\over \partial K^\prime_A}
+i\hbar \left. {\partial \ln J^{1\over 2}\over \partial  g }\right|_{\Phi,K}.
\label{sigmaprimosue2}
\end{equation}
Now, ${\partial \tilde\Sigma\over \partial g}
(\Phi^\prime(\Phi,K),K^\prime (\Phi,K))$ is the same as
$\widetilde{\partial \Sigma\over \partial  g }$.
Let us define
\begin{equation}
S(\Phi,K^\prime, g )={\partial F(\Phi,K^\prime, g )
\over \partial  g }.
\end{equation}
(\ref{differentials}) gives
\begin{equation}
\left. {\partial S\over \partial \Phi^{\prime A}}\right|_{K^\prime}=
\left. {\partial_r \Phi^B\over \partial \Phi^{\prime A}}\right|_{K^\prime}
{\partial S\over \partial \Phi^B}=
{\partial S\over \partial \Phi^B}{(M^{-1})^B}_A.
\end{equation}
Moreover, (\ref{cantrasf}) gives
\begin{equation}
\left. {\partial S\over \partial \Phi^{\prime A}}\right|_{K^\prime}=
{\partial K_B\over \partial  g }{(M^{-1})^B}_A.
\end{equation}
On the other hand, the second of (\ref{differentials})
permits to write
\begin{equation}
\left. {\partial K^\prime_B\over \partial  g }\right|_{\Phi,K}{M^B}_A=
-{\partial K_A\over\partial  g },
\end{equation}
so that we conclude
\begin{equation}
\left. {\partial S\over \partial \Phi^{\prime A}}\right|_{K^\prime}=
-\left. {\partial K^\prime_A\over \partial  g }\right|_{\Phi,K}.
\label{div1}
\end{equation}
Following similar steps, one can prove
\begin{equation}
\left. {\partial S\over \partial K^\prime_A}\right|_{\Phi^\prime}=
{\partial \Phi^{\prime A}\over \partial g}-
{\partial K_B\over \partial g}{(M^{-1})^B}_CF^{CA}
=\left. {\partial \Phi^{\prime A}\over \partial  g }
\right|_{\Phi,K}.
\label{div2}
\end{equation}
This formula, together with
(\ref{div1}) and (\ref{canonical}),
permits to rewrite (\ref{sigmaprimosue2}) in the form
\begin{equation}
{\partial \Sigma^\prime \over \partial  g }=
\widetilde{\partial \Sigma\over \partial  g }
-(S,\tilde\Sigma)+i\hbar
\left. {\partial \ln J^{1\over 2}\over \partial  g }\right|_{\Phi,K}.
\end{equation}
Using (\ref{sigmaprimo}) and (\ref{deltatilde}) we also get
\begin{equation}
{\partial \Sigma^\prime \over \partial  g }=
\widetilde{\partial \Sigma\over \partial  g }
-(S,\Sigma^\prime)+i\hbar \Delta S-i\hbar \left\{
\tilde\Delta S-
\left. {\partial \ln J^{1\over 2}\over \partial  g }\right|_{\Phi,K}\right\}.
\end{equation}
Our thesis will be proved if we are able to show that
\begin{equation}
\tilde\Delta S=
\left. {\partial \ln J^{1\over 2}\over \partial  g }\right|_{\Phi,K}.
\end{equation}
Now, the chain rule gives
\begin{eqnarray}
\tilde\Delta  S&=&(-1)^{\varepsilon_A+1}\left.
{\partial _r\over \partial K^\prime_A}
\left(\left. {\partial S\over \partial\Phi^{\prime A}}\right|_{K^\prime}\right)
\right|_{\Phi^\prime}=(-1)^{\varepsilon_A}\left.
{\partial _r\over \partial K^\prime_A}
\left(\left. {\partial K^\prime _A\over \partial g }\right|_{\Phi,K}\right)
\right|_{\Phi^\prime}\nonumber\\
&=&(-1)^{\varepsilon_A}\left. {\partial_r\over\partial \Phi^B}
\left(\left. {\partial K^\prime _A\over \partial g }\right|_{\Phi,K}\right)
\right|_{K}\left. {\partial_r \Phi^B\over \partial K^\prime_A}
\right|_{\Phi^\prime}+(-1)^{\varepsilon_A}\left. {\partial_r\over\partial K_B}
\left(\left. {\partial K^\prime _A\over \partial g }\right|_{\Phi,K}\right)
\right|_{\Phi}\left. {\partial_r K_B\over \partial K^\prime_A}
\right|_{\Phi^\prime}\nonumber\\
&=&(-1)^{\varepsilon_A}\left. {\partial\over\partial g }\left(
\left. {\partial_rK^\prime_A\over\partial \Phi^B}\right|_K\right)
\right|_{\Phi,K}
\left. {\partial_r \Phi^B\over \partial K^\prime_A}
\right|_{\Phi^\prime}+(-1)^{\varepsilon_A}
\left. {\partial\over\partial g }\left(
\left. {\partial_r K^\prime _A\over\partial K_B}\right|_\Phi\right)
\right|_{\Phi,K}
\left. {\partial_r K_B\over \partial K^\prime_A}
\right|_{\Phi^\prime}.\nonumber\\
\end{eqnarray}
With the help of (\ref{differentials}), one can prove that
\begin{equation}
\matrix{
\left. {\partial_r K^\prime _A\over\partial K_B}
\right|_\Phi={(N^{-1})_A}^B,&
\left. {\partial_r K_B\over \partial K^\prime_A}
\right|_{\Phi^\prime}={N_B}^A-F_{BC}{(M^{-1})^C}_DF^{DA},\cr
\left. {\partial_rK^\prime_A\over\partial \Phi^B}
\right|_K=-{(N^{-1})_A}^CF_{CB},&
\left. {\partial_r \Phi^B\over \partial K^\prime_A}
\right|_{\Phi^\prime}=-{(M^{-1})^B}_CF^{CA}.}
\end{equation}
Moreover, noticing that
\begin{equation}
J^{1\over 2}=\det M^{-1},
\end{equation}
and that
\begin{equation}
d\ln J^{1\over 2}=(-1)^{\varepsilon_B+1}d{N_B}^D{(N^{-1})_D}^B,
\end{equation}
we conclude
\begin{equation}
\tilde\Delta S=\left. {\partial \ln J^{1\over 2}\over \partial  g }\right|_
{\Phi,K}+(-1)^{\varepsilon_B+1}{(M^{-1})^B}_CF^{CA}{(N^{-1})_A}^D
\left. {\partial F_{DB}\over \partial  g }\right|_{\Phi,K}.
\end{equation}
The last term vanishes due to symmetry properties and the statistics
of the various factors. This concludes the proof.

\end{document}